\documentclass[twocolumn]{aastex62}

\newcommand{\pkg}[1]{\texttt{#1}}

\accepted{April 5, 2018}

\shorttitle{GPI Spectra of HR 8799 c,d,e}
\shortauthors{Greenbaum et al.}

\begin{document}

\title{GPI spectra of HR 8799 \MakeLowercase{c}, \MakeLowercase{d}, and
       \MakeLowercase{e} from 1.5 to 2.4$\mu m$ with KLIP Forward Modeling}

\correspondingauthor{Alexandra Greenbaum}
\email{azgreenb@umich.edu}

\author[0000-0002-7162-8036]{Alexandra Z. Greenbaum}
\affiliation{Department of Astronomy, University of Michigan, Ann Arbor, MI 48109, USA}

\author{Laurent Pueyo}
\affiliation{Space Telescope Science Institute, Baltimore, MD 21218, USA}

\author[0000-0003-2233-4821]{Jean-Baptiste Ruffio}
\affiliation{Kavli Institute for Particle Astrophysics and Cosmology, Stanford University, Stanford, CA 94305, USA}

\author[0000-0003-0774-6502]{Jason J. Wang}
\affiliation{Department of Astronomy, University of California, Berkeley, CA 94720, USA}

\author[0000-0002-4918-0247]{Robert J. De Rosa}
\affiliation{Department of Astronomy, University of California, Berkeley, CA 94720, USA}

\author{Jonathan Aguilar}
\affiliation{Department of Physics and Astronomy, Johns Hopkins University, Baltimore, MD 21218, USA}

\author[0000-0003-0029-0258]{Julien Rameau}
\affiliation{Institut de Recherche sur les Exoplan{\`e}tes, D{\'e}partement de Physique, Universit{\'e} de Montr{\'e}al, Montr{\'e}al QC, H3C 3J7, Canada}

\author[0000-0002-7129-3002]{Travis Barman}
\affiliation{Lunar and Planetary Laboratory, University of Arizona, Tucson AZ 85721, USA}

\author[0000-0002-4164-4182]{Christian Marois}
\affiliation{National Research Council of Canada Herzberg, 5071 West Saanich Rd, Victoria, BC, V9E 2E7, Canada}
\affiliation{University of Victoria, 3800 Finnerty Rd, Victoria, BC, V8P 5C2, Canada}

\author[0000-0002-5251-2943]{Mark S. Marley}
\affiliation{NASA Ames Research Center, Mountain View, CA 94035, USA}

\author[0000-0002-9936-6285]{Quinn Konopacky}
\affiliation{Center for Astrophysics and Space Science, University of California San Diego, La Jolla, CA 92093, USA}

\author[0000-0002-9246-5467]{Abhijith Rajan}
\affiliation{School of Earth and Space Exploration, Arizona State University, PO Box 871404, Tempe, AZ 85287, USA}

\author[0000-0003-1212-7538]{Bruce Macintosh}
\affiliation{Kavli Institute for Particle Astrophysics and Cosmology, Stanford University, Stanford, CA 94305, USA}

\author[0000-0003-4142-9842]{Megan Ansdell}
\affiliation{Department of Astronomy, University of California, Berkeley, CA 94720, USA}

\author[0000-0001-6364-2834]{Pauline Arriaga}
\affiliation{Department of Physics \& Astronomy, University of California, Los Angeles, CA 90095, USA}

\author[0000-0002-5407-2806]{Vanessa P. Bailey}
\affiliation{Jet Propulsion Laboratory, California Institute of Technology, Pasadena, CA 91109, USA}

\author{Joanna Bulger}
\affiliation{Subaru Telescope, NAOJ, 650 North A{'o}hoku Place, Hilo, HI 96720, USA}

\author[0000-0002-3099-5024]{Adam S. Burrows}
\affiliation{Department of Astrophysical Sciences, Princeton University, Princeton NJ 08544, USA}

\author[0000-0001-6305-7272]{Jeffrey Chilcote}
\affiliation{Kavli Institute for Particle Astrophysics and Cosmology, Stanford University, Stanford, CA 94305, USA}
\affiliation{Department of Physics, University of Notre Dame, 225 Nieuwland Science Hall, Notre Dame, IN, 46556, USA}

\author[0000-0003-0156-3019]{Tara Cotten}
\affiliation{Department of Physics and Astronomy, University of Georgia, Athens, GA 30602, USA}

\author{Rene Doyon}
\affiliation{Institut de Recherche sur les Exoplan{\`e}tes, D{\'e}partement de Physique, Universit{\'e} de Montr{\'e}al, Montr{\'e}al QC, H3C 3J7, Canada}

\author[0000-0002-5092-6464]{Gaspard Duch\^ene}
\affiliation{Department of Astronomy, University of California, Berkeley, CA 94720, USA}
\affiliation{Univ. Grenoble Alpes/CNRS, IPAG, F-38000 Grenoble, France}

\author[0000-0002-0176-8973]{Michael P. Fitzgerald}
\affiliation{Department of Physics \& Astronomy, University of California, Los Angeles, CA 90095, USA}

\author[0000-0002-7821-0695]{Katherine B. Follette}
\affiliation{Physics and Astronomy Department, Amherst College, 21 Merrill Science Drive, Amherst, MA 01002, USA}

\author[0000-0003-3978-9195]{Benjamin Gerard}
\affiliation{University of Victoria, 3800 Finnerty Rd, Victoria, BC, V8P 5C2, Canada}
\affiliation{National Research Council of Canada Herzberg, 5071 West Saanich Rd, Victoria, BC, V9E 2E7, Canada}

\author[0000-0002-4144-5116]{Stephen J. Goodsell}
\affiliation{Gemini Observatory, 670 N. A'ohoku Place, Hilo, HI 96720, USA}

\author{James R. Graham}
\affiliation{Department of Astronomy, University of California, Berkeley, CA 94720, USA}

\author[0000-0003-3726-5494]{Pascale Hibon}
\affiliation{Gemini Observatory, Casilla 603, La Serena, Chile}

\author[0000-0003-1498-6088]{Li-Wei Hung}
\affiliation{Department of Physics \& Astronomy, University of California, Los Angeles, CA 90095, USA}

\author{Patrick Ingraham}
\affiliation{Large Synoptic Survey Telescope, 950N Cherry Ave., Tucson, AZ 85719, USA}

\author{Paul Kalas}
\affiliation{Department of Astronomy, University of California, Berkeley, CA 94720, USA}
\affiliation{SETI Institute, Carl Sagan Center, 189 Bernardo Ave.,  Mountain View CA 94043, USA}

\author{James E. Larkin}
\affiliation{Department of Physics \& Astronomy, University of California, Los Angeles, CA 90095, USA}

\author{J\'er\^ome Maire}
\affiliation{Center for Astrophysics and Space Science, University of California San Diego, La Jolla, CA 92093, USA}

\author[0000-0001-7016-7277]{Franck Marchis}
\affiliation{SETI Institute, Carl Sagan Center, 189 Bernardo Ave.,  Mountain View CA 94043, USA}

\author[0000-0003-3050-8203]{Stanimir Metchev}
\affiliation{Department of Physics and Astronomy, Centre for Planetary Science and Exploration, The University of Western Ontario, London, ON N6A 3K7, Canada}
\affiliation{Department of Physics and Astronomy, Stony Brook University, Stony Brook, NY 11794-3800, USA}

\author[0000-0001-6205-9233]{Maxwell A. Millar-Blanchaer}
\affiliation{Jet Propulsion Laboratory, California Institute of Technology, Pasadena, CA 91109, USA}
\affiliation{NASA Hubble Fellow}

\author[0000-0001-6975-9056]{Eric L. Nielsen}
\affiliation{SETI Institute, Carl Sagan Center, 189 Bernardo Ave.,  Mountain View CA 94043, USA}
\affiliation{Kavli Institute for Particle Astrophysics and Cosmology, Stanford University, Stanford, CA 94305, USA}

\author{Andrew Norton}
\affiliation{University of California Observatories/Lick Observatory, University of California, Santa Cruz, CA 95064, USA}

\author[0000-0001-7130-7681]{Rebecca Oppenheimer}
\affiliation{Department of Astrophysics, American Museum of Natural History, New York, NY 10024, USA}

\author{David Palmer}
\affiliation{Lawrence Livermore National Laboratory, Livermore, CA 94551, USA}

\author{Jennifer Patience}
\affiliation{School of Earth and Space Exploration, Arizona State University, PO Box 871404, Tempe, AZ 85287, USA}

\author[0000-0002-3191-8151]{Marshall D. Perrin}
\affiliation{Space Telescope Science Institute, Baltimore, MD 21218, USA}

\author{Lisa Poyneer}
\affiliation{Lawrence Livermore National Laboratory, Livermore, CA 94551, USA}

\author[0000-0002-9667-2244]{Fredrik T. Rantakyr\"o}
\affiliation{Gemini Observatory, Casilla 603, La Serena, Chile}

\author[0000-0002-8711-7206]{Dmitry Savransky}
\affiliation{Sibley School of Mechanical and Aerospace Engineering, Cornell University, Ithaca, NY 14853, USA}

\author{Adam C. Schneider}
\affiliation{School of Earth and Space Exploration, Arizona State University, PO Box 871404, Tempe, AZ 85287, USA}

\author[0000-0003-1251-4124]{Anand Sivaramakrishnan}
\affiliation{Space Telescope Science Institute, Baltimore, MD 21218, USA}

\author[0000-0002-5815-7372]{Inseok Song}
\affiliation{Department of Physics and Astronomy, University of Georgia, Athens, GA 30602, USA}

\author[0000-0003-2753-2819]{Remi Soummer}
\affiliation{Space Telescope Science Institute, Baltimore, MD 21218, USA}

\author{Sandrine Thomas}
\affiliation{Large Synoptic Survey Telescope, 950N Cherry Ave., Tucson, AZ 85719, USA}

\author{J. Kent Wallace}
\affiliation{Jet Propulsion Laboratory, California Institute of Technology, Pasadena, CA 91109, USA}

\author[0000-0002-4479-8291]{Kimberly Ward-Duong}
\affiliation{Physics and Astronomy Department, Amherst College, 21 Merrill Science Drive, Amherst, MA 01002, USA}

\author{Sloane Wiktorowicz}
\affiliation{Department of Astronomy, UC Santa Cruz, 1156 High St., Santa Cruz, CA 95064, USA }

\author[0000-0002-9977-8255]{Schuyler Wolff}
\affiliation{Leiden Observatory, Leiden University, P.O. Box 9513, 2300 RA Leiden, The Netherlands}

\begin{abstract}

We explore KLIP forward modeling spectral extraction on Gemini Planet Imager
coronagraphic data of HR 8799, using \pkg{PyKLIP} and show algorithm
stability with varying KLIP parameters. We report new and re-reduced
spectrophotometry of HR 8799 c, d, and e in H \& K bands. We discuss a strategy
for choosing optimal KLIP PSF subtraction parameters by injecting simulated
sources and recovering them over a range of parameters. The K1/K2 spectra for
HR 8799 c and d are similar to previously published results from the same
dataset. We also present a K band spectrum of HR 8799 e for the first time and
show that our H-band spectra agree well with previously published spectra from
the VLT/SPHERE instrument. We show that HR 8799 c and d show significant
differences in their H \& K spectra, but do not find any conclusive differences
between d and e or c and e, likely due to large error bars in the recovered
spectrum of e.  Compared to M, L, and T-type field brown dwarfs, all three
planets are most consistent with mid and late L spectral types. All objects are
consistent with low gravity but a lack of standard spectra for low gravity
limit the ability to fit the best spectral type. We discuss how dedicated
modeling efforts can better fit HR 8799 planets' near-IR flux and discuss how
differences between the properties of these planets can be further explored.

\end{abstract}


\section{Introduction} \label{sec:intro}

Directly imaged planets present excellent laboratories to study the properties
of the outer-architectures of young solar systems. Near-infrared spectroscopic
follow-up can constrain atmospheric properties including molecular absorption,
presence of clouds, and non-equilibrium chemistry \citep{barman2011,
konopacky2013, marley2012}.  Composition, especially in relation to the host
star is an important probe of physical processes and formation history
\citep{oberg2011}.

HR 8799 is a 1.5 $M_\odot$ star \citep{graykaye1999, baines2012} located at a
distance of $39.4 \pm 1.0 pc$ \citep{vanleeuwen2007} with an estimated age of
$\sim30 Myr$ \citep{moor2006, marois2008, hinz2010, zuckerman2011, baines2012,
malo2013} based on it's likely membership in the Columba association. It
contains multiple imaged planets b, c, d, and e \citep{marois2008,marois2010}
witthat lie between 10 and 100 AU separations from the host star.
\citet{lavie} show the inner 3 planets fall within the $\mathrm{H_2O}$ and
$\mathrm{CO_2}$ ice lines based on a vertically isothermal, passively
irradiated disk model.  \citet{konopacky2013} point out that in this region
non-stellar C and O abundances are available to accrete onto a planetary core,
and that the measured abundances of HR 8799 planets could help distinguish
between core-accretion scenario and gravitational instability, which is
expected to produce stellar-abundances. 

HR 8799 has been a testbed for detection techniques
\citep[e.g.,][]{lafreniere2009,soummer2011}, astrometric monitoring and
dynamical studies \citep{fabrycky2010,soummer2011,pueyo2015, konopacky2016,
zurlo2016, wertz2017}, atmospheric characterization \citep{janson2010,
bowler2010, hinz2010,barman2011, madhusudhan2011, currie2011, skemer2012,
marley2012, konopacky2013, ingraham2014,barman2015,rajan2015, bonnefoy2016},
and even variability \citep{apai2016}. Studying the properties of multiple
planets in the same system presents a unique opportunity for understanding its
formation, by studying dynamics and composition as a function of mass and
semi-major axis. 

Spectrophotometry and moderate resolution spectroscopy have provided a
detailed view into the atmospheres of the HR 8799 companions. Water and carbon
monoxide absorption lines have been detected in the atmospheres of planets b
and c, with methane absorption additionally detected in b \citep{barman2011,
konopacky2013, barman2015}.  To account for the discrepancy between the spectra
of b and c and those of field brown dwarfs, various studies based on near-IR
observations from 1-5$\mu$m have proposed the presence of clouds
\citep[e.g.][]{marois2008, hinz2010, madhusudhan2011}, disequilibrium chemistry
to explain an absence of methane absorption \citep{barman2011, konopacky2013},
and non-solar composition \citep{lee2013}.  However, some work suggests
that prescriptions of disequilibrium chemistry, non-solar composition, and/or
patchy atmospheres may not play an important role for the d, e planets
\citep[e.g.][]{bonnefoy2016}, which appear consistent with dusty late-L objects
based on their YJH spectra and SEDs, and can be modeled with atmospheres that
do not contain these features. 
K band spectra are especially sensitive to atmospheric properties and
composition and can probe the presence of methane and water. HR 8799 b, c, and
d have shown a lack of strong methane absorption in the K-band spectra
\citep{bowler2010, barman2011, currie2011, konopacky2013, ingraham2014},
inconsistent with field T-type brown dwarfs. 

Stellar PSF subtraction algorithms that take advantage of angular and/or
spectral diversity, while powerful for removing the stellar PSF, result in
self-subtraction of the signal of interest, which can bias the measured
astrometry and photometry \citep{marois2010spie}.  Self-subtraction biases in
the signal extraction are commonly avoided by injecting negative simulated
planets in the data and optimizing over the residuals
\citep[e.g.,][]{lagrange2010}. For multi-wavelength IFU data, template PSFs of
representative spectral types can be used to optimize the extraction/detection
\citep{marois2014spie, gerard2016spie, ruffio2017}. An analytic forward model
of the perturbation of the companion PSF due to self-subtraction effects can be
a more efficient approach that is less dependent on the template PSF and
algorithm parameters \citep{pueyo2016}.

In this paper we present H \& K spectra of HR 8799 c, d, and e obtained with
the Gemini Planet Imager (GPI). We use Karhunen Lo\'eve Image Projection (KLIP)
for PSF subtraction \citep{soummer2011} with the forward model formalism
demonstrated in \cite{pueyo2016}. In \S \ref{sec:obs} we describe our
observations and data reduction. This is followed by a brief description of
KLIP forward modeling (hereafter KLIP-FM) and discussion of the stability of
our extracted signal with varying KLIP parameters. In \S \ref{sec:results} we
present our recovered spectra alongside previous results
\citep{oppenheimer2013, ingraham2014, zurlo2016, bonnefoy2016}, and discuss
consistencies and discrepancies. We also describe a method to calculate the
similarity of the three extracted spectra and discuss our findings.  We compare
a library of classified field and low gravity brown dwarfs to our H and K
spectra in \S \ref{sec:comparison} and report best fit spectral types.
Finally, we discuss a few different atmospheric models and their best fits to
our spectra in \S \ref{sec:models}.  We summarize and discuss these results in
\S \ref{sec:conclusions}. In Appendix \ref{sec:residuals} we show our detail
residuals between the processed data and the forward models. Appendix
\ref{sec:algos} we compare the forward model extraction described in this study
with other algorithms. In Appendix \ref{sec:chi2_by_band} we show planet
comparisons by individual bands. We provide our spectra in Appendix
\ref{sec:spectrum}.

\section{Observations and Data Reduction \label{sec:obs}} \label{sec:observations}

\subsection{GPI Observations and Datacube Assembly}

\begin{table}[b]
\caption{Summary of observations \label{tab:obs}}
\begin{center}
\begin{tabular}{l|c|c|c|c|c|c}
     DATE & Band & $N_{frame}$ & $\mathrm{t_{frame}}$ & $\Delta \mathrm{PA}$
& Airmass & Seeing  \\ \hline\hline
    2013/11/17 & K1 & 24 & 90 $s$ & $17.1^o$ & 1.62 & $0\farcs98$ \\
    2013/11/18 & K2 & 20 & 90 $s$ & $9.7^o$ & 1.62 & $0\farcs72$ \\
    2016/09/19 & H & 60 & 60 $s$ & $20.9^o$ & 1.61 & $0\farcs97$ \\
\end{tabular}
\end{center}
\end{table} 

HR 8799 was observed with the Gemini Planet Imager Integral Field Spectrograph
(IFS) \citep{macintosh2014} with its K1 and K2 filters on 2013 November 17
(median seeing $0\farcs97$) and November 18 (median seeing 0.75 arcsec),
respectively, during GPI's first light.  The data were acquired with a
continuous field of view (FOV) rotation near the meridian transit to achieve
maximum FOV rotation suitable for ADI processing \citep{marois2006}. Conditions
are described in detail in \cite{ingraham2014}. During the last 10 exposures of
the K1 observations cryocooler power was decreased to 30\% to reduce vibration,
and the last 14 exposures of the K2 observations had the cryocooler power
decreased. Since commissioning linear-quadratic-Gaussian control has been
implemented \citep{poyneer2016} and the cryocooling system has been upgraded
with active dampers to mitigate cryocooler cycle vibrations. HR 8799 was
observed again on September 19 2016 in GPI's H-band (median seeing $0\farcs97$)
as a part of the GPI Expolanet Survey (GPIES) with the updated active damping
system. Planet b falls outside the field of view in these data. Table
\ref{tab:obs} summarizes all GPI observations of HR 8799 used in this study. 

Datacube assembly was performed using the GPI Data Reduction Pipeline (DRP)
\citep{gpipipeline,perrin2014, perrin2016}. Wavelength calibration for the K1
and K2 data was done using a Xenon arc lamp and flexure offset adjusted
manually \citep{wolff2014}.  Bad pixels were corrected and dark and sky frames
were subtracted from the raw data. The raw detector frames were assembled into
spectral datacubes. Images were corrected for distortion \citep{konopackyspie}
and high pass filtered. H-band datacube reduction followed a similar procedure,
except that flexure offsets were automatically determined based on
contemporaneous arc lamp images. \citet{wang2018} contains a thorough description
of standard data reduction procedures. 

The instrument transmission function was calibrated using the GPI grid apodizer
spots, which place a copy of the stellar PSF in four locations in the image
\citep{sivarmakrishnan2006, marois2006spots}. These fiducial satellite spots
are used to convert raw data counts to contrast units and to register and
demagnify the image \citep{maire2014, wang2014}. 

\subsection{KLIP forward modeling for unbiased spectra} \label{sec:klip}

Stellar PSF subtraction is performed by constructing an optimized combination
of reference images with KLIP \citep[For a complete description
see][]{soummer2012}. Reference images are assembled from the
full dataset to take advantage of both angular and spectral diversity. The KL
basis, $Z_{k}$ is formed from the covariance matrix of the reference images and
projected onto the data $I(n)$ to subtract the Stellar PSF:
\begin{eqnarray} \label{eqn:klip}
S = \sum(I(n) - \sum_{k=1}^{k_{klip}}<I, Z_{k}>_S Z_{k}(n)) \end{eqnarray}
where $S$ is the klipped data and $Z_{k}$ is determined by the reference
library selection criteria. To account for over- and self- subtraction of the
companion signal we use the approach detailed in \cite{pueyo2016} to forward
model the signal in PSF-subtracted data to recover an unbiased spectrum. The
forward model is constructed by perturbing the covariance matrix of the
reference library to account for a faint companion signal and propagating this
through the KLIP algorithm  for additional terms, as we show in
Equation \ref{eqn:klipfm}. Over- and self-subtraction effects are accounted
for in the forward model by projecting a model of the PSF,
$F_{model}$, onto the unperturbed KL basis, $Z_{k}$, (over-subtraction) and
projecting the PSF model onto the the KL basis perturbation (self-subtraction),
where the KL basis perturbation, $\Delta Z_{k}$, is a function of the
unperturbed KL basis, $Z_k$, and the PSF model, $F_{model}$.  The forward model
is constructed by the terms that are linear in the planet signal:
\begin{eqnarray} \label{eqn:klipfm}
FM &=& \sum F_{model} \nonumber \\ 
&-& \sum < F_{model},Z_{k}> Z_{k}  \nonumber \\ 
&-& \sum <Z_k , Z_k> \Delta Z_k 
 - \sum <Z_k, \Delta Z_k> Z_k 
\end{eqnarray}
For the analysis presented in this paper, the PSF model, $F_{model}$, is
constructed from the satellite spots. 

\begin{figure}
\centering
\includegraphics[width=3.4in]{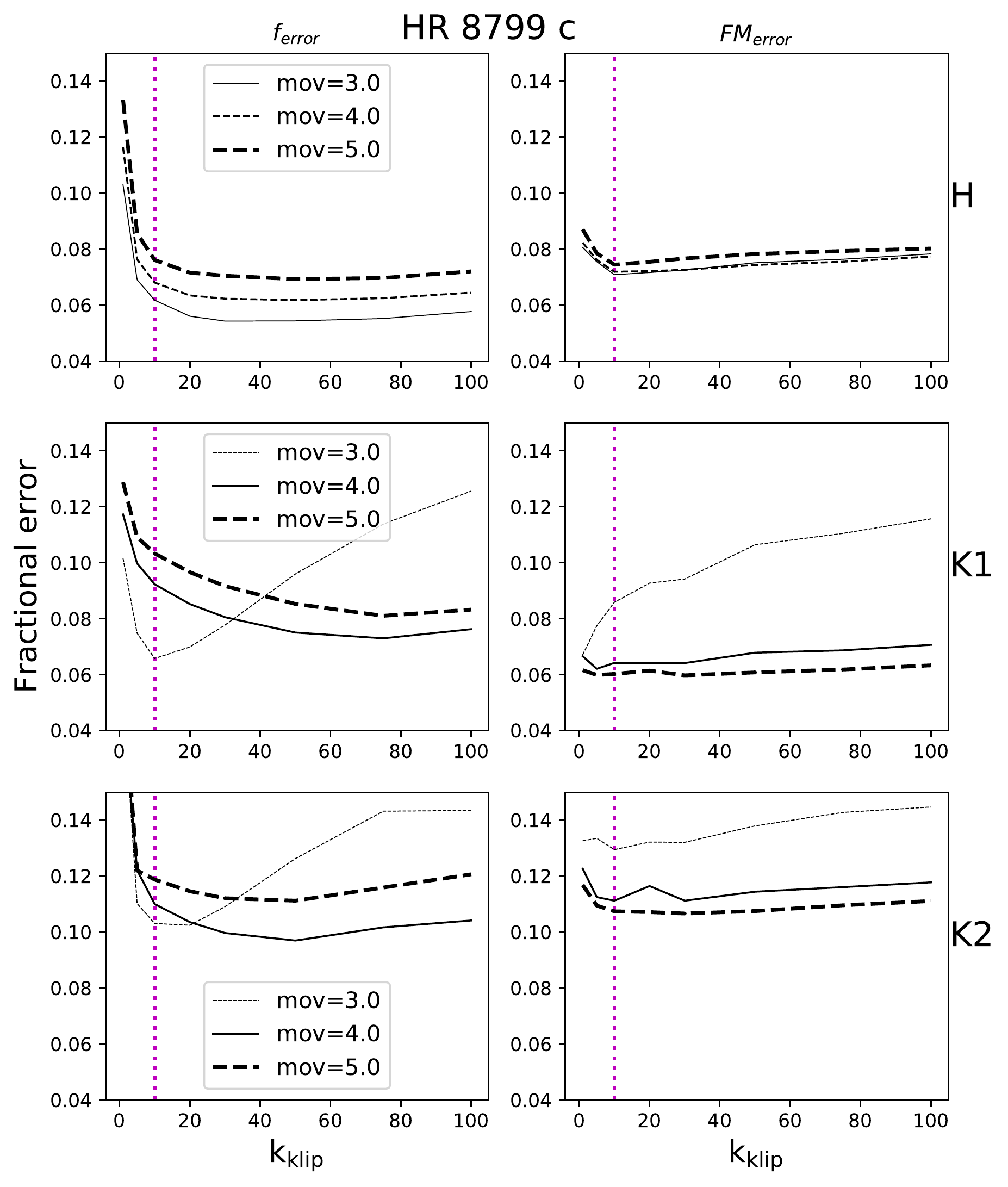} \caption{Residual
errors over a range of KLIP-FM parameters for HR 8799 c. Left: Difference
between the injected spectrum at the planet location and the recovered spectra
of simulated injections, normalized by the recovered spectrum (Equation
\ref{eqn:spect_error}).  Right: The residual error of the forward model
normalized by the sum of the pixels. (Equation \ref{eqn:fm_error}). The solid
line denotes the chosen exclusion criterion, or $mov$ value (in pixels). The
number of KL components used, $k_{klip}$, denoted by the vertical dotted line,
is chosen in a region where $f_{error}$ is decreasing at the selected value of
$mov$. } 
\label{fig:bias_c}
\end{figure}

\begin{figure}
    \centering
    \includegraphics[width=3.4in]{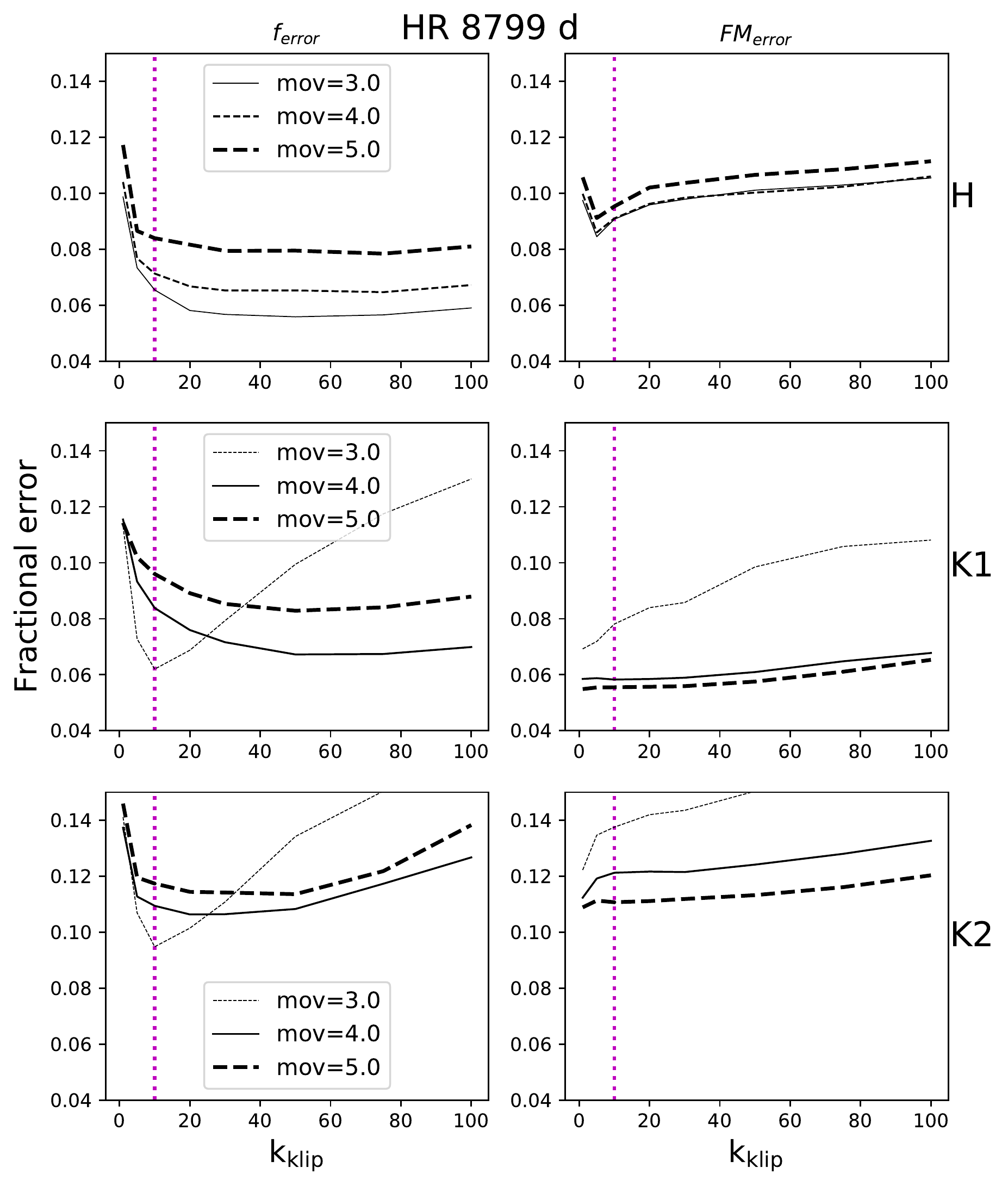}
    \caption{Same as Figure \ref{fig:bias_c}, but for HR 8799 d. The intersection
of the solid curve and verticle dotted line denote our choice of parameters.}
    \label{fig:bias_d}
\end{figure}

\begin{figure}
    \centering
    \includegraphics[width=3.4in]{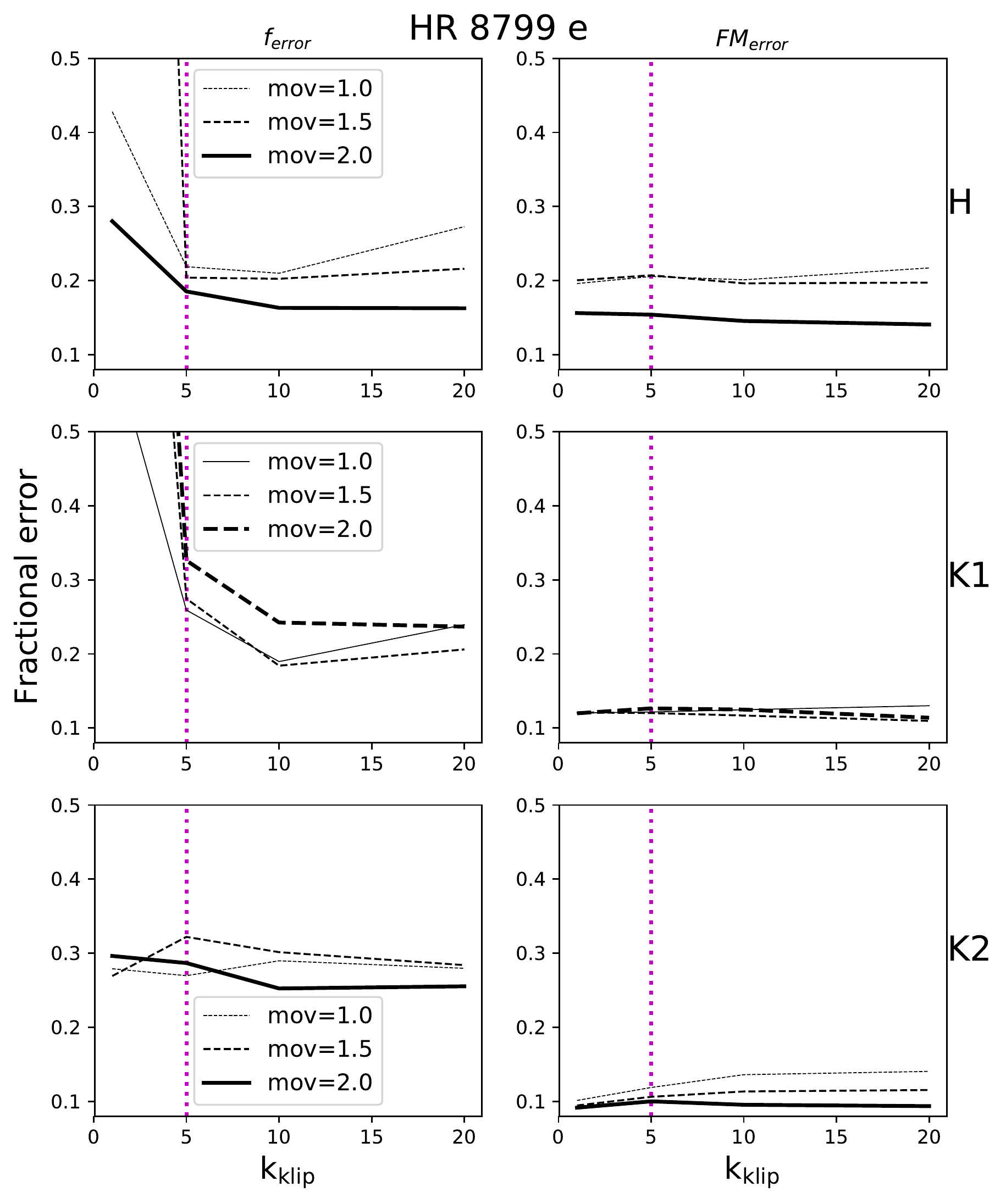}
    \caption{Same as Figure \ref{fig:bias_c}, but for HR 8799 e. The intersection
of the solid curve and verticle dotted line denote our choice of parameters.}
    \label{fig:bias_e}
\end{figure}

After computing the forward model, the spectrum is recovered by solving the
inverse problem for $f_{\lambda}$:

\begin{eqnarray} \label{eqn:axeqb}
f_{\lambda}\cdot FM = S
\end{eqnarray}
where $S$ is the stellar PSF-subtracted data processed with KLIP, as in
Equation \ref{eqn:klip}. This assumes the relative astrometry
has already been calculated.
Using the approximate contrast summed over the bandpass for each object we run
Bayesian KLIP Astrometry \citep{wang2016} to measure the astrometry of each
planet in the different datasets first. The improved position reduces residuals
between the forward model containing the optimized spectrum and the
PSF-subtracted data.  The procedures and documentation are available in the
\pkg{PyKLIP}\footnote{http://pyklip.readthedocs.io/} package.

PCA-based PSF subtraction, especially which includes the signal of the
companion as in the case of ADI and Simultaneous Spectral Differential Imaging
(SSDI), will bias the extracted spectrum \citep{marois2006spie, pueyodloci}.
This bias is often seen as a sensitivity to algorithm parameters. We run KLIP
Forward Modeling (KLIP-FM) spectral extraction considering the
effect of two KLIP parameters, the KLIP cuttoff $k_{klip}$, which sets the
number of KL modes used for the subtraction, and movement (or aggressiveness),
$mov$, which defines the maximum allowed level of overlap between the planets
position in a given image and other images selected for its reference library.
See \citet{ruffio2017} for a more formal definition.  We vary these parameters
as a proxy for understanding how biased and noisy our extraction is. As in
\cite{pueyo2016} we expect this forward modeling approach to be less sensitive
to changing algorithm parameters than regular PCA-style subtractions, and
overall this is what we observe. However, there is still some 2nd-order
dependence either due to the model being wrong or noise in the image. 

We examine how the spectral extraction results vary with algorithm parameters
through two measures of error, error in the spectrum extraction $f_{error}$ and
residual error around the location of the signal $FM_{error}$. To measure error
in the spectral extraction, artificial signals are inserted into the
data. We simulate 11 artificial signals evenly distributed (30
deg apart) in an annulus at the same separation but avoiding the position angle
of the planet.  The artificial sources are simulated with spectra corresponding
to the spectrum measured from the planet with KLIP-FM.  We define $f_{error}$
as 
\begin{eqnarray} \label{eqn:spect_error}
f_{error}= \frac{1}{N_{\lambda}}\sum_{\lambda}^{N_\lambda}
\sqrt{\frac{\sum_i^{N_{sim}}(f_{\lambda} -
f'_{\lambda,i})^2}{N_{sim}f_\lambda^2}}
\end{eqnarray} 
where $f_\lambda$ is the spectrum recovered at the location of the planet and
$f'_{\lambda,i}$ is the spectrum recovered for the $i$th artificial source of
total $N_{sim}$ sources. Our spectral datacubes contain $N_\lambda = 37$
wavelength slices per band. We define the residual error, $FM_{error}$, as the
square root of the sum of the residual image pixels squared divided by the sum
of the klipped image of the planet squared. The residual is calculated inside a
region with a radius of 4 pixels centered on the planet.  
\begin{eqnarray} \label{eqn:fm_error}
R &=&\sum_{\lambda}^{N_{\lambda}} S_\lambda - FM_\lambda, \nonumber \\
FM_{error}&=& \sqrt{\frac{\sum_{pix} R^2} {\sum_{pix}{S^2}}} 
\end{eqnarray} Appendix \ref{sec:residuals} contains the full residuals in the
region around each planet for each band.

Figures \ref{fig:bias_c}, \ref{fig:bias_d}, and \ref{fig:bias_e}, for planets
c, d, and e respectively, display the spectrum error (left) and residual error
(right) for the range of investigated KLIP parameters. 
We see that the error converges as $k_{klip}$ increases as demonstrated in
\citet{pueyo2016}, when the model is capturing the signal. In certain cases
when the model is wrong, the error does not converge as we see for planets c
and d K-band data when $mov=3$.
These metrics show the stability of of the forward model solution with KLIP
parameters. The solid line plotted in each panel represents the value of $mov$
chosen for the final spectrum, with the other values of $mov$ represented in
dashed lines of varying thickness. The dotted vertical magenta line represents
the chosen value of $k_{klip}$.  For planet e we excluded the
two closest simulated sources to avoid contamination from the real planet
signal.

We choose KLIP parameters that minimize the $f_{error}$ term and prefer
solutions with smaller value of $k_{klip}$ that occur before the minimum. We
also check that the residual error $FM_{error}$ stays relatively flat. As
demonstrated in \cite{pueyo2016} the forward model starts to fail for larger
$k_{klip}$ when the signal is bright.  The error in the residual is generally
close to the spectrum error, $f_{error}$, except in the K1 and
K2 reductions of e, when the spectrum error could be reflecting more residual
speckle noise. 
The spectrum error term is fairly well behaved for all three planets, in
general flattening with $k_{klip}$. The H band data, for which our reduction
show the most stable behavior with KLIP parameters, has more rotation and was
taken after several upgrades to the instrument.   The behavior of the two error
metrics for HR 8799 e improved when wavelength slices from the band edges were
removed prior to PSF subtraction. 

\begin{figure*}[htbp!]
    \centering
    \includegraphics[width=5in]{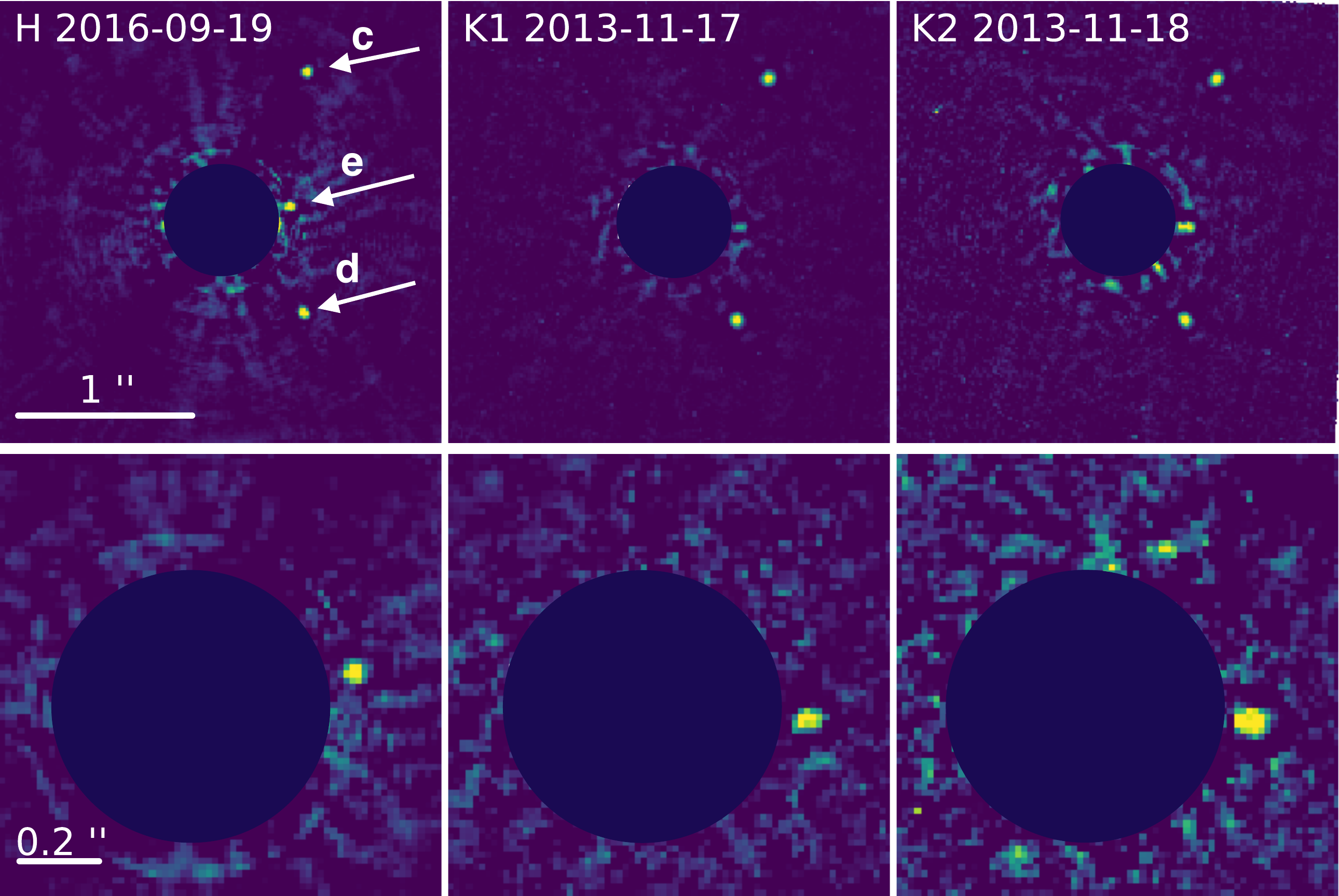}
    \caption{\textbf{Top:} Our standard KLIP subtracted cubes with $mov$ = 3
pixels. \textbf{Bottom:} Subtracted cubes with $mov$ = 1 pixel. The data are
zoomed in to highlight HR 8799. The H band data quality is higher due to larger 
field rotation. We see the relative position difference of
planet e due to orbital motion between the 2013 (K band) and 2016 (H band)
epochs.}
    \label{fig:fullcubes}
\end{figure*}

For HR 8799 c and d we note that most of the parameter combinations yield a
similar level of error, within a few percent. Changing the klip parameters near
our chosen values does not have a large effect on the spectrum. For HR 8799 e
the bias is generally higher (note the scale in Figure \ref{fig:bias_e}). This
is reflected in the larger error bars for e reported in our final spectrum. We
display our collapsed datacubes reduced through \pkg{PyKLIP} in Figure
\ref{fig:fullcubes} showing a less aggressive reduction (larger $mov$) used to
extract spectra of HR 8799 c and d in the top panel, and more aggressive
reduction (smaller $mov$, including more images in the reference library) used
to extract the spectrum of HR 8799 e in the bottom row.

\section{Results using optimized KLIP parameters} \label{sec:results}

After inspecting an initial reduction with all data, we remove slices at the
ends of each cube where the signal is low (due to low filter
throughput at band edges) and re-rerun our reduction. We take this step to
avoid biasing the spectrum extracted in Equation \ref{eqn:axeqb} with datacube
slices that contain no signal.  KLIP errors are computed from the standard
deviation of the simulated source recovery at each wavelength channel. 
Errorbars displayed reflect the standard deviation of the spectrum recovered
from simulated sources and the uncertainty in the satellite spot flux,
calculated from measuring the variation of the spot flux photometry in this
data.

\begin{figure*}
    \centering
    \includegraphics[width=5in]{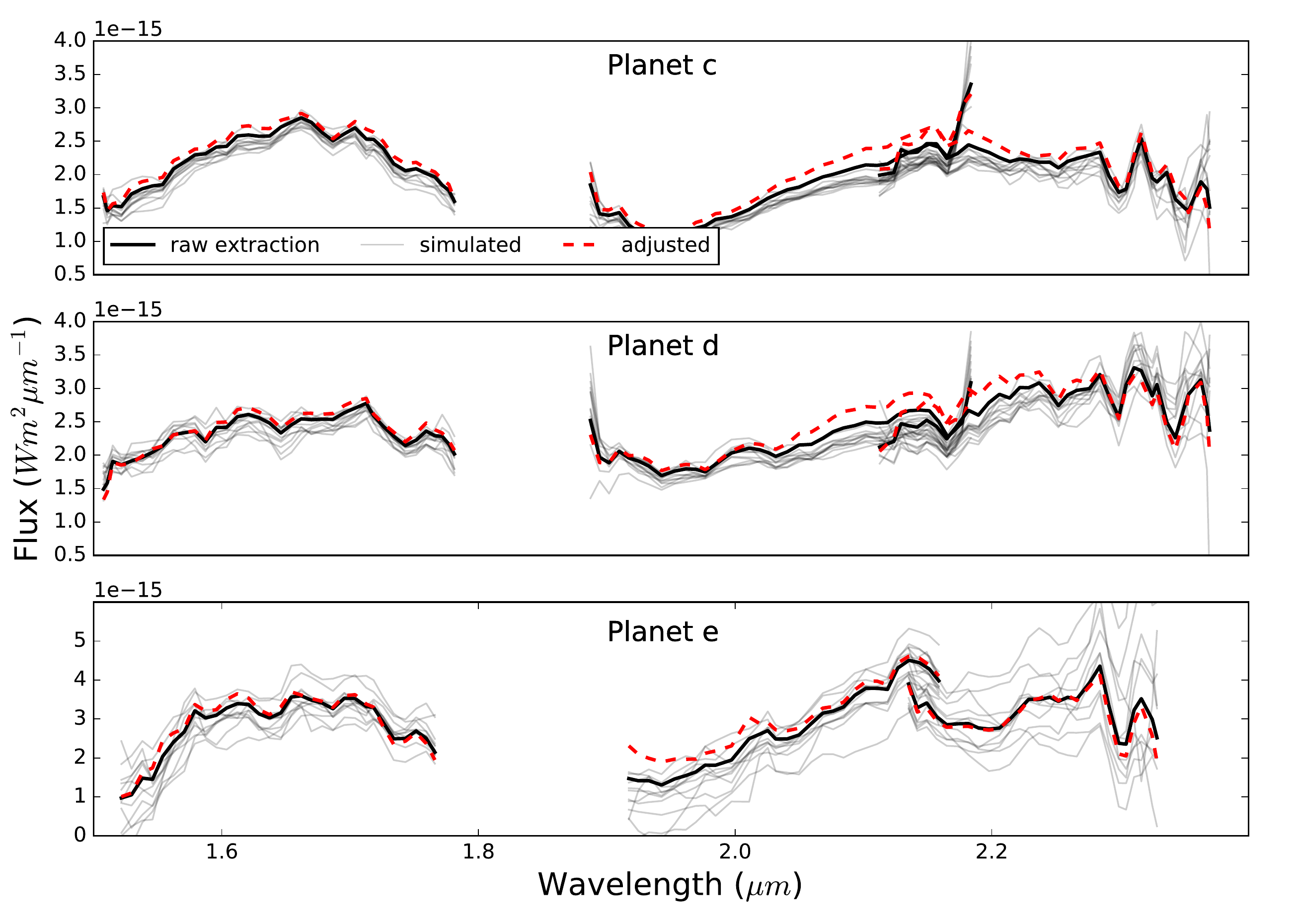}
    \caption{Resulting flux at 10pc of KLIP-FM spectrum extraction for each
planet (solid black line) and recovery of artificial sources (gray lines),
simulated with the matching spectrum. Increases in flux at the edges of the K1
band, where the filter throughput is low, are not significant. These are not
seen in the case of e since these datta were processes without the wavelength
slices at the edge of the band. The red dotted line shows the adjusted spectrum
applying the flux loss factor in Equation \ref{eqn:adjust}.}
    \label{fig:fakebias}
\end{figure*}

We find that in some parts of the spectrum, the scatter in flux of the
recovered injected signals is not symmetric about the injected spectrum, which
indicates that the model is slightly wrong to a scaling factor.  This effect is
typically on the order of $5-10\%$ and is most obvious between $2.0-2.2\mu m$
for c and d, and at the short wavelength edge of K1 for e. In Figure
\ref{fig:fakebias} we show the the recovered spectrum and the recovered
artifical sources.  To account for bias in spectral extraction, we have
adjusted the spectrum  by a scaling term that accounts for the flux loss, which
is the factor between the recovered spectrum and the mean spectrum recovered
from the simulated sources, 
\begin{eqnarray} \label{eqn:adjust} F_\lambda = \frac{f_{\lambda,k}}{\sum
f_{sim\lambda,k} /N_{sim}}
\end{eqnarray}
for the $\mathrm{k_{th}}$ planet at each wavelength slice,
$\lambda$. Where the scatter is more symmetric (such as in all the H-band
datasets) the model is correctly accounting for the flux and the adjusted
spectrum matches the initial reduction. All following figures and calculations
in this paper use the adjusted spectrum.

\begin{figure*}
    \centering
    \includegraphics[width=7in]{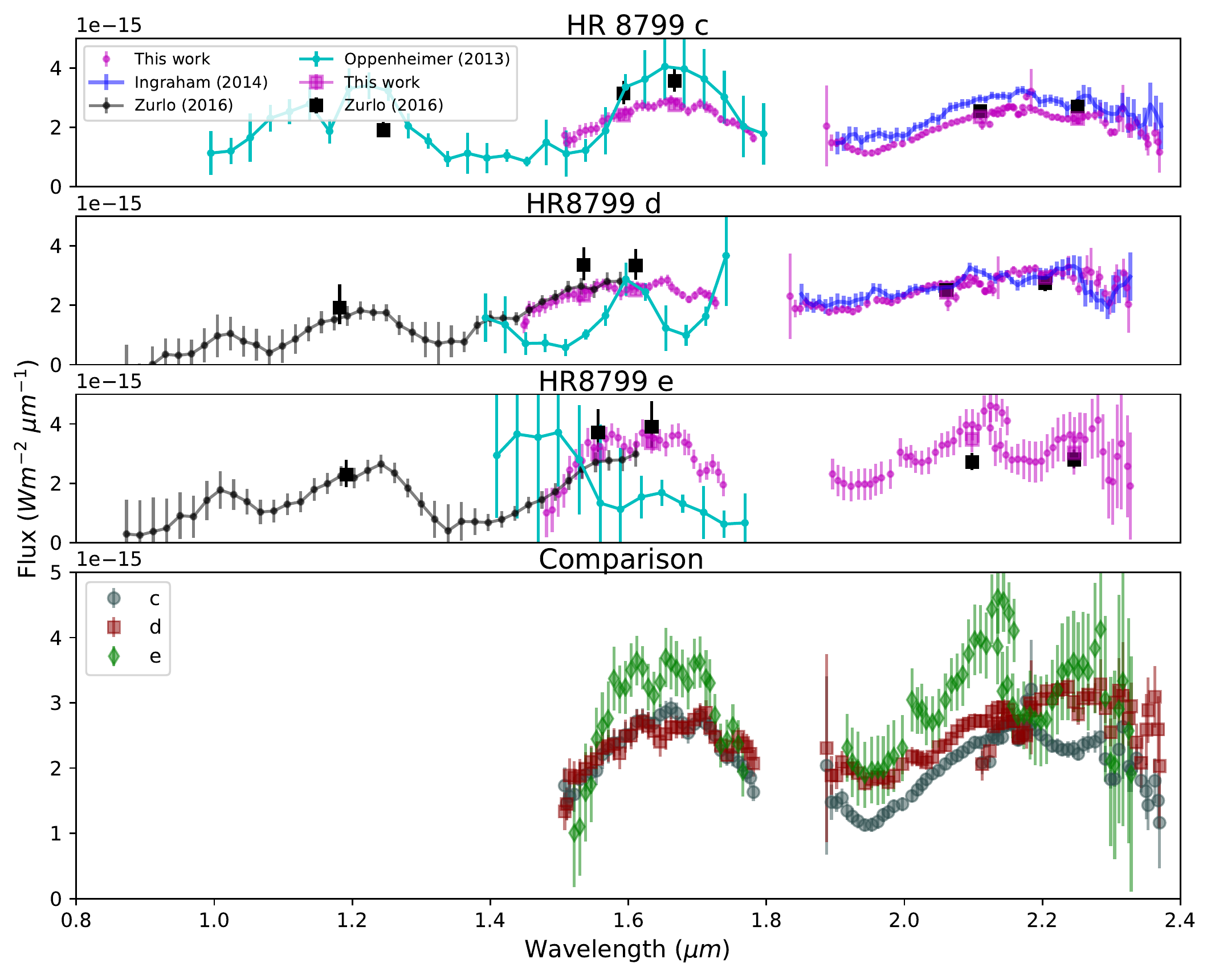}
    \caption{Spectra recovered with KLIP FM for HR 8799 c, d, and e showing
flux at 10pc. Overplotted are the original GPI K-band spectra of c and d for
the same dataset in dark blue \citep{ingraham2014}, YJH spectra from
Palomar/P1640 \citep{oppenheimer2013} (KLIP reduction) in cyan
points, YJH spectra and JHK photometry using the VLT/SPHERE
instrument\citep{zurlo2016}  in black points and squares. The P1640 points are
scaled from normalized flux units to match our data for this comparison. The
bottom panel shows a comparison of our three recovered spectra. }
    \label{fig:spectra_flux}
\end{figure*}

As in \cite{bonnefoy2016}, we use a Kurucz spectrum at 7500K, matched to the
photometry of HR 8799 A, to convert contrast to flux. We display results for
the best KLIP parameters in Figure \ref{fig:spectra_flux}, adjusting our
spectrum for c as indicated in Figure \ref{fig:fakebias}. Cyan points represent
Palomar/P1640 data from \cite{oppenheimer2013}, which are scaled from
normalized flux to match the rest of the points plotted. Dark blue lines for c
and d panels are the K band spectra from the same dataset previous published in
\cite{ingraham2014}. In black are SPHERE/IFS YJH spectra published by
\cite{zurlo2016} and blue squares show SPHERE/IRDIS photometry.

Our K-band spectrum for HR 8799 e changed the most with varying KLIP parameters.
We note a discrepancy between the overlapping edges of our K1 and K2 spectra
for e. This is unlikely to be a calibration error since it is not seen in the
cases of the c and d spectra (except at the very edges of the band where the
transmission is very low. Based both on photometry from \cite{zurlo2016} and
our residual errors (see Appendix \ref{sec:residuals}) the K1 fluxes may not be
correct.  We suspect the K2 reduction is more representative of the true
spectrum. We note that our K2 spectrum of e more closely resembles that of d
than K1. 

We find very similar morphology as the previously published spectra for c and
d, although slightly lower flux in the case of c. Since these planets are so
bright we may still be over-subtracting, if the linear approximation in the
forward model is not appropriate, as described in \cite{pueyo2016}. Our results
are consistent with SPHERE/IRDIS K1 \& K2 photometery within errorbars, but
systematically lower. \cite{ingraham2014} noted the K-band spectra in
particular, combined with photometry at 3 and 4 $\mu$m showed a lack of methane
absorption, and our re-reduction is consistent. They noted the flatter spectrum
for d, which also appears to be the case for our new H-band spectra compared
with c and e.

The SPHERE IRDIS H-band photometry are discrepant with our result. However,
these photometry are also discrepant with the SPHERE IFS spectra. Our KLIP-FM
H-band spectra for d and e are in good agreement with those obtained from the
SPHERE IFS, within error bars. Towards the center of H band we see a slight dip
in the spectrum for HR 8799 e, between 1.6 and 1.7 $\mu$m, which is not seen in
the SPHERE IFS spectrum. However, there due to correlated noise, this may not be 
a real effect. Y and J observations of HR 8799 with GPI will improve
the comparison and provide a complete YJHK spectra on the same instrument.

\subsection{Comparison of c, d, and e} \label{sec:chi2}

Differences between the three spectra could show evidence of varying
atmospheric composition and formation histories, or first order
physical effects such as clouds, temperature, and gravity.  In the bottom
panel of Figure \ref{fig:spectra_flux} we plot all three spectra on the same
axes. The H band spectrum of e appears to be most discrepant from the other
two, and there are differences between all three in K1-K2. We note that our K1
and K2 spectra for e do not match in the overlap region around 2.18 $\mu m$.
This is only the case for e, which could indicate that is it not due to the
algorithm or flux calibration, but more speckle residuals close in. The
residual images in Appendix \ref{sec:residuals} also show a possible speckle
influencing the forward model solution for K1 data. The short wavelength edge
of K2 suggests the spectrum is more similar to that of d. 

These small differences motivate a more quantitative comparison. We compute
$\chi^{2}_{i,j}$ between each spectrum and a spectrum drawn randomly from its
error:
\begin{eqnarray}
\label{eqn:chi2hist}
\chi^{2}_{i,j} = \frac{1}{N_\lambda -1}(f_i - f_j^*)^T Cov_i^{-1} (f_i - f_j^*)
\end{eqnarray}
where $f$ is the spectrum of the $i$th object normalized by its sum, and
$f_j^*$ is a spectrum drawn randomly from the sum-normalized spectrum of the
$j$th assuming Gaussian errors, taking into account covariance $Cov_j$ (where
the errors are scaled by the same normalization factor). $Cov_i$ and $Cov_j$
are the covariance matrices of the $i$th and $j$th planets computed as
described in \cite{grecobrandt2016} and \cite{derosa2016}.  Here we draw
$f_j^*$ and compute this statistic $10^5$ times and compare the resulting
distributions for each planet.  We consider the full H-K spectrum so that
relative flux between bands is preserved.

\begin{figure}
    \centering
    \includegraphics[width=3in]{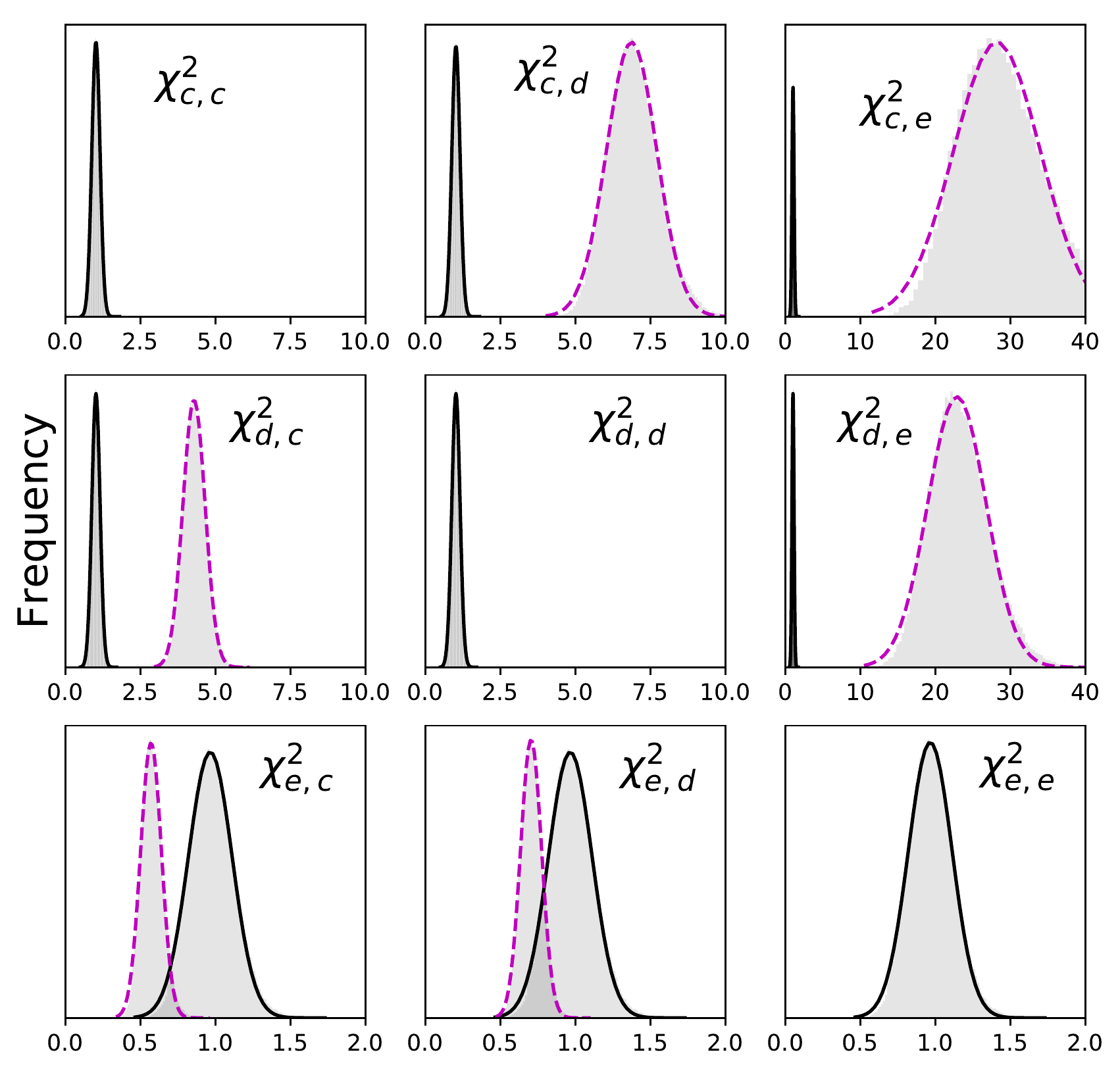}
    \caption{A cross-comparison of all three planets showing the distribution
of $\chi^2_{i,j}$ (defined in Equation \ref{eqn:chi2hist}). The diagonal shows
each spectrum compared with $10^5$ random draws from itself within the error
bars. These, in solid outline are repeated for each panel in the same row. The
off-diagonal panels also show the comparison with random draws from a different
object spectrum, as indicated. These show clear ($>5\sigma$) discrepancy between
c and d planets.}
    \label{fig:hists}
\end{figure}

We plot the histograms of $\chi^{2}$ in Figure \ref{fig:hists}.  We compare the
$\chi^{2}$ distribution of each spectrum with one drawn randomly from the same
spectrum ($i=j$ case, diagonals), with the $\chi^2$ distribution of each
spectrum compared to one drawn randomly from the other two ($i\ne j$ case,
off-diagonals).  The results show a discrepancy between c and d to $> 5\sigma$.
There is a less significant discrepancy between HR 8799 c and e and between d
and e. While the $\chi^2$ distributions of c-e and c-c and the distributions of
d-e and d-d appear distinct, there is still some overlap in the $\chi^2$
distributions of e-e and e-c and between e-e and e-d. This lack of symmetry is
likely due to larger errorbars of the HR 8799 e spectrum. These results show a
discrepancy between e and c to $\sim1.8\sigma$ and between e and d to
$\sim1.2\sigma$. Reducing the errors for the planet e spectrum could improve
this comparison.  Resolving the discrepancy in the spectrum of e between K1 and
K2 bands edges should also improve this comparison.

We repeated the same comparison for each of H, K1, and K2 bands separately,
where the spectra and errors are normalized within each band.  We show the
detailed results in Appendix \ref{sec:extrahistograms}. In this case we do not
find the same differences between the spectra. This indicates that the relative
level of flux between bands is the dominant effect.

\section{Comparison to field objects} \label{sec:comparison}
We compare our results with known field objects as described in
\cite{chilcote2017}. We compare our H \& K spectrophotometry with a library of
$\sim1600$ spectra for M,L and T-dwarf field objects. These are compiled from
the SpeX Prism library \citep{burgasser2014}, the IRTF Spectral Library
\citep{cushing2005}, the Montreal Spectral Library
\citep{gagne2015,robert2016}, and from \cite{allersliu2013}. Each spectrum and
uncertainty was binned to the spectral resolution of GPI ($R\sim45-80$
increasing from H to K2). We convolve the spectrum with a Gaussian function of
width matching the resolution for that band. The uncertainties are normalized
by the effective number of spectral channels within the convolution width.
Spectra that are incomplete in the GPI filter coverage are excluded from the
fit. 

Spectral type classifications were obtained from various literature sources,
specified for individual objects.  For objects that had both optical and
near-IR spectral types, the near-IR spectral type was used. 
Gravity classifications were assigned from the literature as either old field
dwarfs ($\alpha$, FLD-G), intermediate surface gravity ($\beta$, INT-G), or low
surface gravity, such as typically seen in young brown dwarfs ($\gamma$,
$\delta$). 
Several studies \citep{kirkpatrick2005,kirkpatrick2006, cruz2009} outline the
$\alpha$, $\beta$, $\gamma$ classification scheme, including an additional
$\delta$ classification from \cite{kirkpatrick2005} to account for even lower
gravity features, based on optical spectra. FLD-G, INT-G, VL-G, based on
Near-IR spectra, follows \citep{allersliu2013}.  The results between the two
classification schemes are correlated \citep[as discussed in][]{allersliu2013}
but do not always match.

\begin{figure}
    \centering
    \includegraphics[width=3.25in]{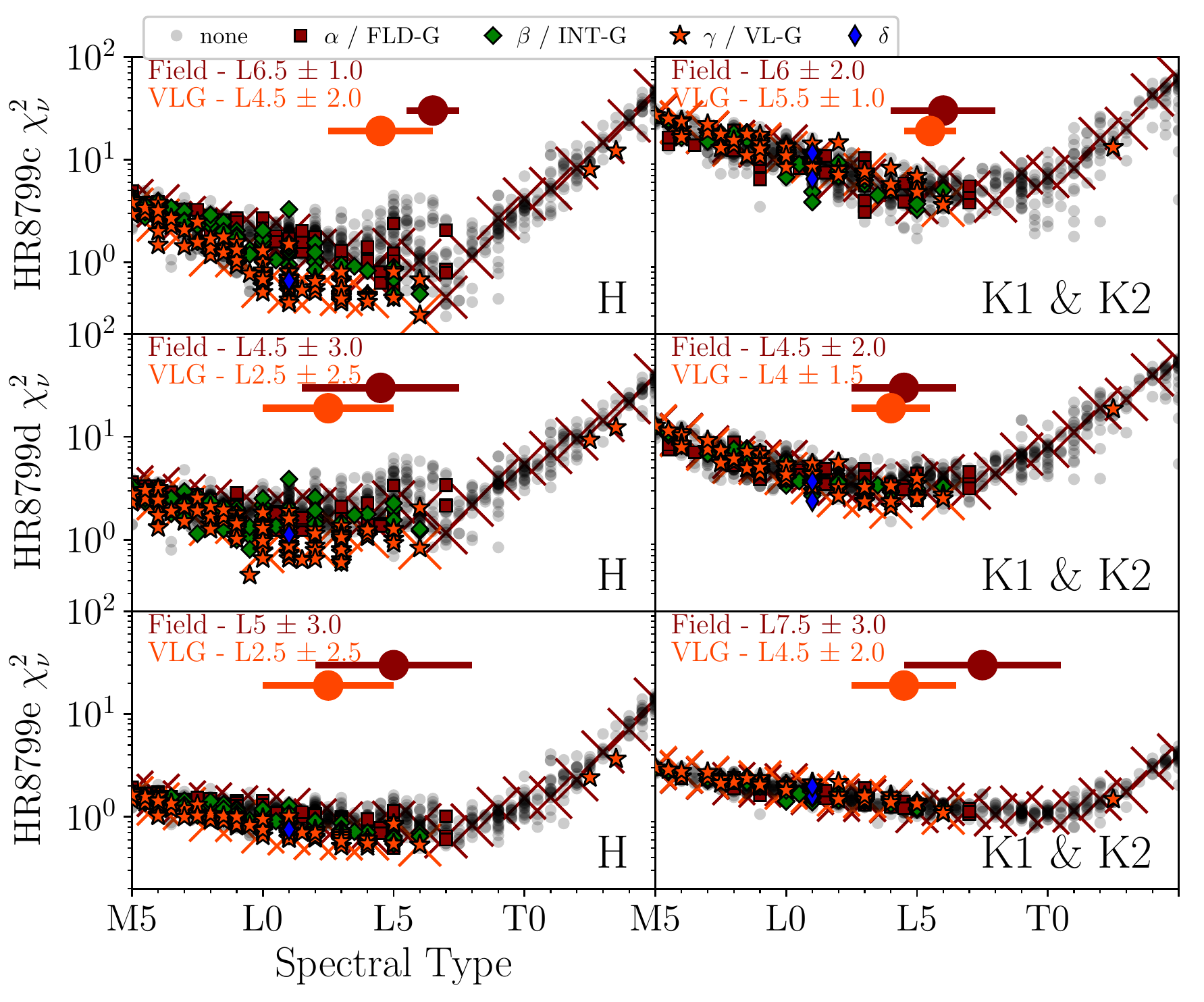}
    \caption{We plot $\chi^2_\nu$  between our GPI spectrum and each object in
the spectral libraries described, as a function of spectral type of field
objects for each planet. From top to bottom we plot planets c, d, and e. The
left shows $\chi^2$ for H band, the right shows the $\chi^2$ for combined K1,
K2 bands. The large red and orange points at the top of each panel represent
the mean and $1-\sigma$ error for the best fit SED for field and VLG objects,
respectively. We indicate gravity classification by the legend at the top.
Spectral standards for FLD-G \citep{burgasser2014,kirkpatrick2010} and VL-G
\citep{allersliu2013} are indicated by red and yellow crosses, respectively.}
    \label{fig:sedfit_bands}
\end{figure}

First we compare our spectra with those of each object in the compiled library,
separately for H and K1-K2 bands. We compute reduced $\chi^2$ using the binned
spectra of comparison objects. The results for each of HR 8799 c, d, and e are
shown in Figure \ref{fig:sedfit_bands}. Spectral standards are marked for
gravity classification where the classification is known. 

\begin{figure*}
    \centering
    \includegraphics[scale=0.5]{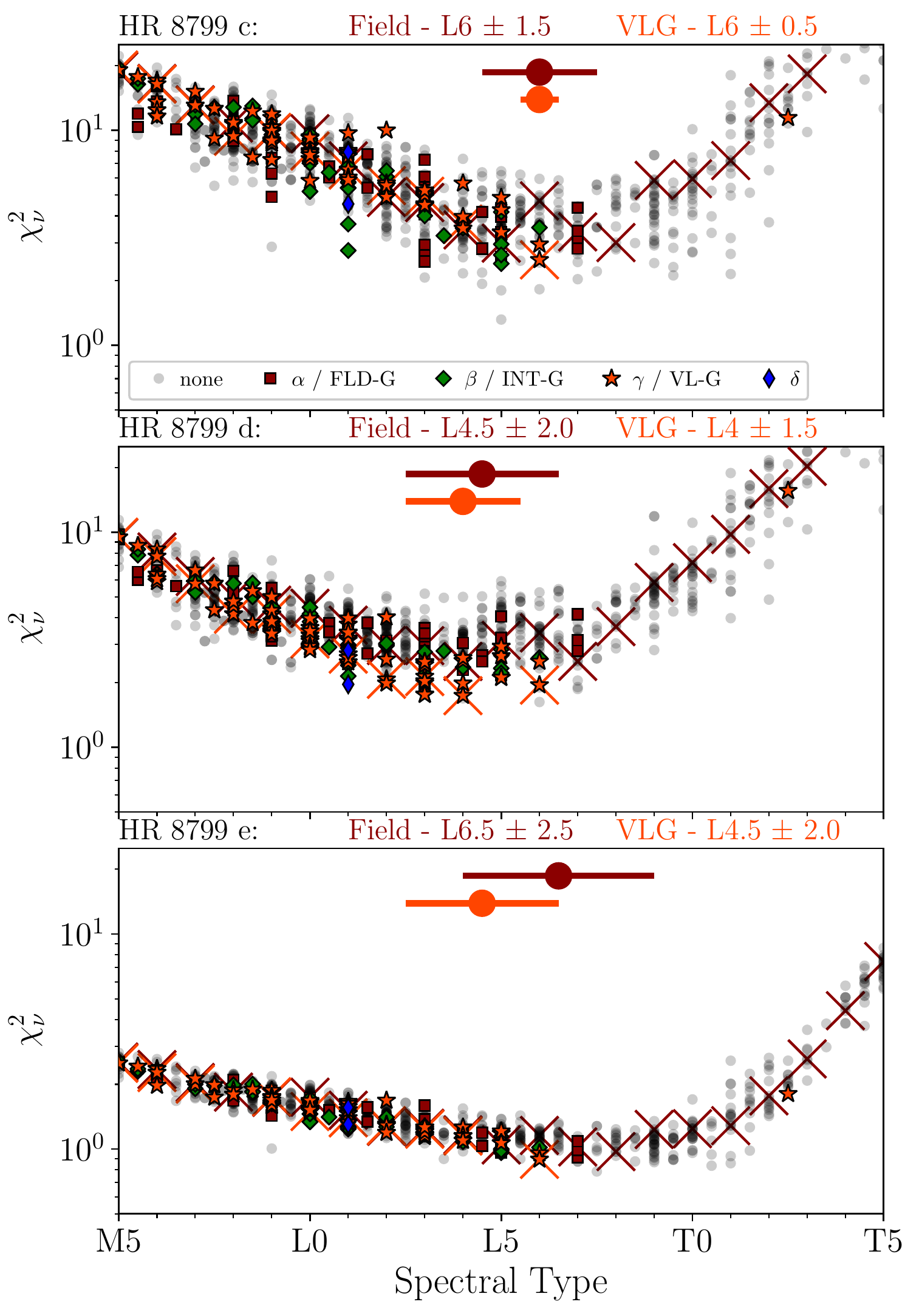}
    \includegraphics[scale=0.5]{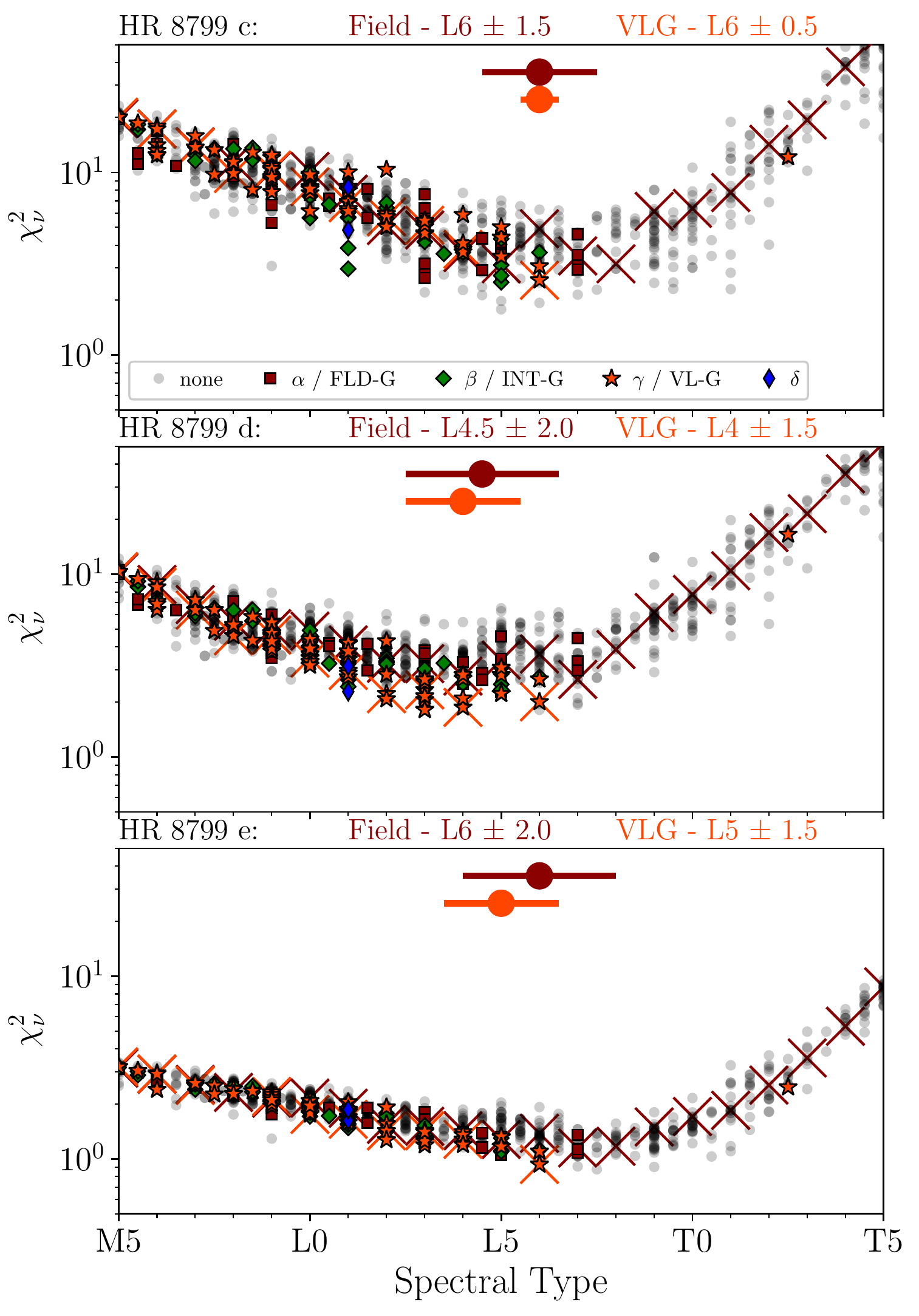}
    \caption{\textbf{Left:} \textit{Unrestricted} $\chi^2$ fit of spectral
library objects to the combined H \& K GPI spectrum. The unrestricted fit
allows the normalization to vary between H and K1+K2 bands.  \textbf{Right:}
\textit{Restricted} $\chi^2$ fit of spectral library objects to the combined H
\& K GPI spectrum. The restricted fit only allows the normalization to vary
within the uncertainty of photometric calibration. The two agree within error
in all cases.} \label{fig:sedfit}
\end{figure*}

Next we simultaneously fit both H, K1, and K2 bands by computing $\chi^2$
between the spectrum of each object in these libraries and our GPI spectra, in
an \textit{unrestricted} and \textit{restricted} fit. The unrestricted fit is
done with  independent normalization between  bands and summing $\chi^2$ for
each band, shown in the left panel of Figure \ref{fig:sedfit}.  For the
restricted fit the normalization can only vary within the uncertainty in the
photometric calibration \citep{maire2014}. The restricted fit is displayed in
the right panel of  Figure \ref{fig:sedfit}. The definition of $\chi^2$ in the
restricted fit is described in \cite{chilcote2017}; we repeat it here for
clarity: 

$\chi^2$ comparison between each of our spectra and the $k$th object in the
library is defined as follows: 
\begin{eqnarray} 
\chi^2_k =&& \sum^2_{j=0}\sum^{n_j}_{i=0}
\Bigg[\frac{F_j(\lambda_i)-\alpha_k\beta_{j,k}C_{j,k}(\lambda_i)}{\sqrt{\sigma^2_{F_j}(\lambda_i)
+\sigma^2_{C_{j,k}}(\lambda_i)}}\Bigg]^2 \nonumber\\ &&+
\sum^2_{j=0}\Bigg[\frac{\beta_{j,k}-1}{\sigma_{m_j}}\Bigg],
\end{eqnarray} summed over bands, $j$ and $n_j$ wavelength channels in each
band. $F_j(\lambda_i)$ and $\sigma_{F_j}(\lambda_i)$ are the measured flux and
uncertainty in the $j$th band and $i$th wavelength channel. $C_{j,k}$ and
$\sigma_{C_{j,k}}(\lambda_i)$ are the corresponding binned flux and uncertainty
of the $k$th object. $\alpha_k$ is a scale factor that is the same for each
band and $\beta_{j,k}$ is a band-dependent scale factor, chosen to minimize
this term. The first term represents the unrestricted $\chi^2$, where each band
can vary freely by scaling factor $\beta_{j,k}$. The second cost term compares
$\beta_{j,k}$ to the satellite fractional spot flux uncertainty measured in
each band, $\sigma_{m_{j}}$ \citep{maire2014}. 

\begin{figure}
    \centering
    \includegraphics[width=3.3in]{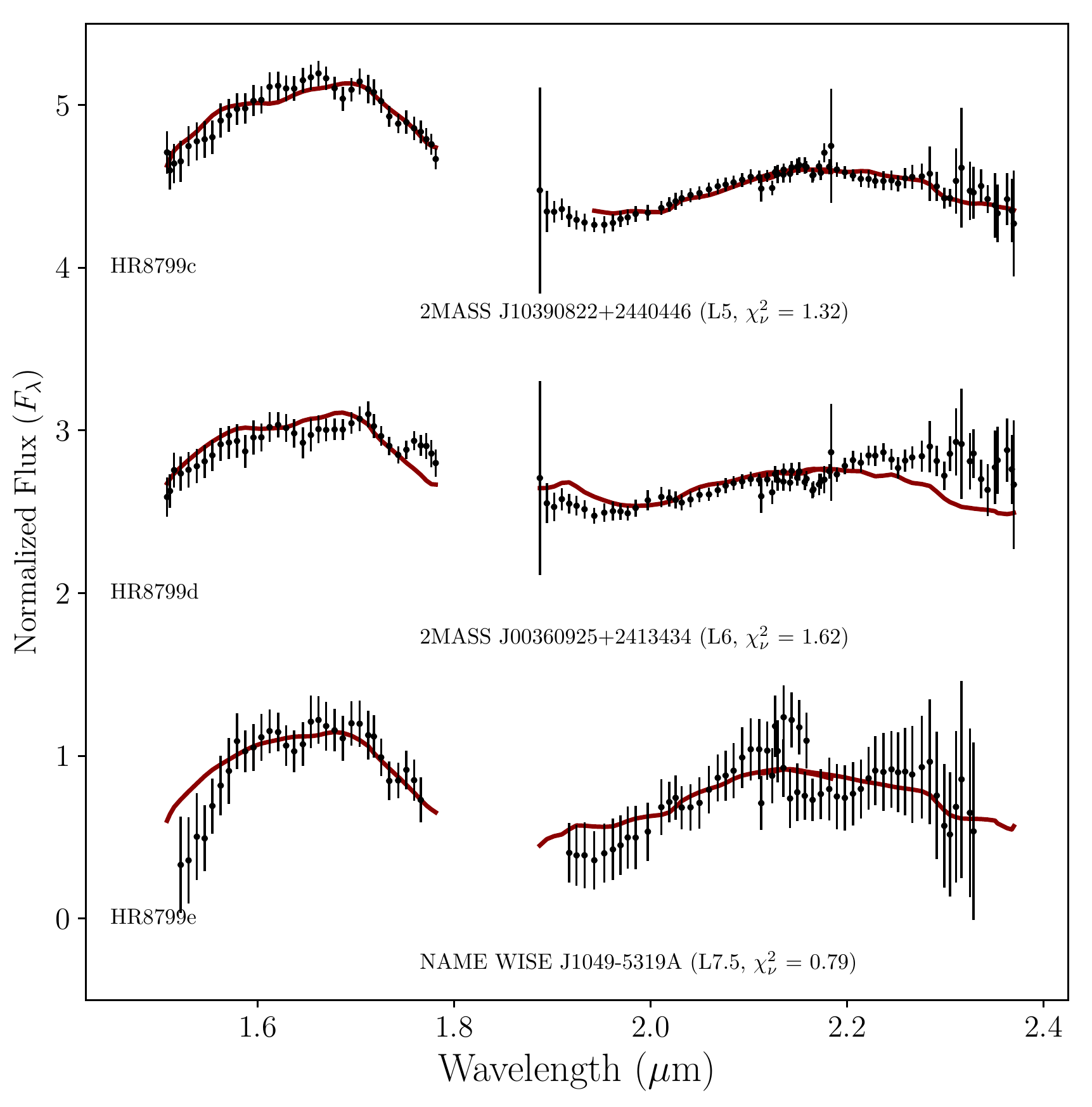}
    \caption{Best fit object to each of HR 8799 c, d, and e spectra within the
described spectral library. Both the unrestricted and restricted case fits
yielded the same best fitting object spectrum \citep{zhang2009, chiu2006,
schneider2014, luhman2013}.}
    \label{fig:bestfit}
\end{figure}

Lastly, we show the best fit object spectrum from our spectral library
overplotted on the GPI spectra in Figure \ref{fig:bestfit}. We show these for
both unrestricted and restricted cases. The object names,  spectral types, and
reduced $\chi^2$ are displayed. 

In general, each object is best represented by a mid-to-late L-type spectrum. A
lack of spectral standards for gravity indicators for late L-types limits the
gravity classification based on these fits. The unrestricted and restricted
fits generally agree for spectral type. For the unrestricted fit the same
object provides the best fit for both c and e.

Planet c is consistent with spectral type $\sim$ L6, both for the individual
band fits and the simultaneous fits. The fits to low gravity types indicate
earlier spectral type, but likely due to a lack of spectral standards for late
L to T-type objects. The H-band spectrum fit, in particular, indicates low
gravity (yellow stars). Both unrestricted and restricted fits yield a spectral
type L6.0$\pm$1.5 for planet c. The best fit object for both fits is 2MASS
J10390822+2440446, which has spectral type L5 \citep{zhang2009}. 

In the case of planet d, the H band spectrum is less peaked. Spectral type
$\sim$L4.5 is best fit for both individual bands and simultaneous fits. Again,
the H-band fit tends to favor low gravity. The simultaneous fit gives spectral
type L4.5$\pm$2.0 for field and L4 $\pm$1.5 for VLG objects, for both the
unrestricted and restricted cases. The best fit object in both cases, 2MASS
J00360925+2413434, spectral type L6 \citep{chiu2006,schneider2014}. 

The individual H and K fits are flatter for e. The H-band and K-band individual
fits are consistent with mid-to-late L-type spectrum. The K-band part of the
spectrum is consistent with a wide range of spectral types, extending to early
T, due in part to large error bars.  The simultaneous fit gives spectral type
L6.5$\pm$2.5 for the unrestricted fit and L6$\pm$2.0. Both restricted and
unrestricted cases yield the best fit for WISE J1049-5319A \citep{luhman2013},
classified as type L7.5 \cite{burgasser2013}. 

Better wavelength coverage would improve spectral type fitting, as well as a
larger library of near-IR spectra and photometry for comparison objects from the
field. More low-gravity standards at late spectral types would also also
improve the VLG fits. Resolving the discrepancy at the edge of K1 and K2 would
also help constrain best fit spectral type for planet e. Variability studies
may show additional evidence of cloud holes, a characteristic of objects
between L- and T- spectral types \cite{radigan2014}.

\section{Comparison to model spectra} \label{sec:models}
We compare our HR 8799 c,d,e spectra with several atmospheric models that have
been presented in previous studies to fit the planet spectra and/or photometry.
This section is broken up into two sections. The first is a comparison of our
spectra to best fit atmospheric models from previous work, A PHOENIX model
\cite{barman2011} that provided the best fit to HR 8799 c in
\citet{konopacky2013}, and a set of models from \cite{saumonmarley2008}, which
we refer to as \textit{Patchy Cloud} models, that provided the best fit for HR
8799 c and d in \citet{ingraham2014}. We compare these to highlight differences
between the three spectra and see how well the models hold up to the new data
in H-band. The second section compares our spectra to two model grids with
varying effective temperature and gravity. The two model grids are the
CloudAE-60 model
\cite{madhusudhan2011}\footnote{http://www.astro.princeton.edu/$\sim$burrows/8799/8799.html
} and the BT-Settl model
\cite{baraffe2015}\footnote{https://phoenix.ens-lyon.fr/Grids/BT\-Settl/CIFIST2011\_2015/}.

\begin{figure*}
    \centering
    \includegraphics[scale=0.85]{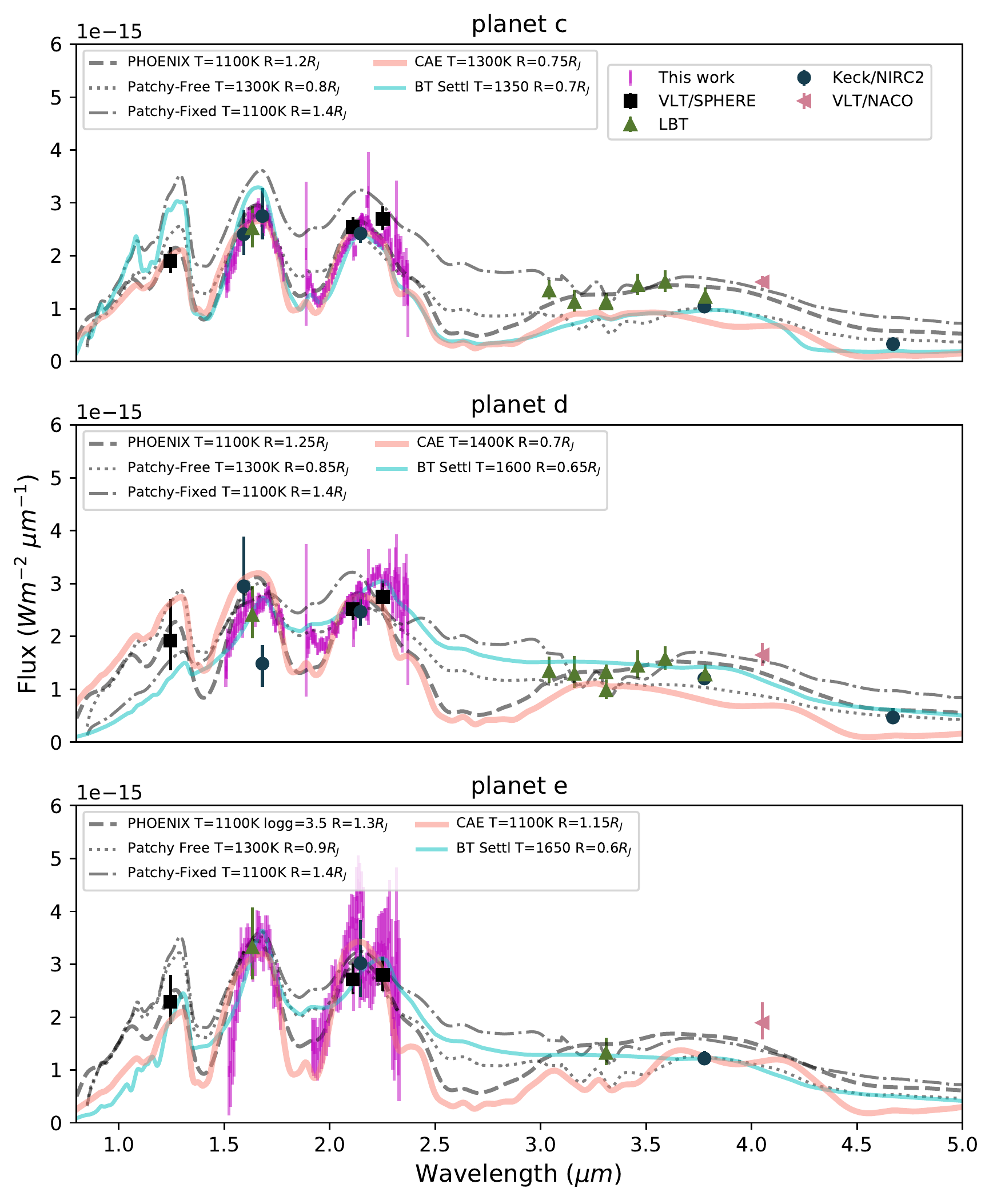}
    \caption{Atmospheric models are plotted in various line styles indicated by
the legend for HR 8799 c (top), d (middle), and e (bottom). GPI spectra are
plotted as magenta bars. Normalized Phoenix models is displayed with a a thick
gray line. For the \cite{saumonmarley2008} patchy cloud models, the normalized
models are plotted in thin solid blue lines, while the fixed models are plotted
in dash-dot lines.  Cloud-AE models are plotted as a dotted line, and BT-Settl
models as dashed lines. We also plot broadband photometry from previous work,
with symbols corresponding to each instrument. Black squares correspond to
VLT/SPHERE IRDIS \citep{zurlo2016}, teal circles to Keck/NIRC2
\citep{marois2008,marois2010,galicher2011,currie2011}, green vertical triangles
to LBT \citep{skemer2012,skemer2014}, and pink left-pointing triangles to NACO
\citep{currie2014}.}
    \label{fig:allmodels}
\end{figure*}

For each set of models, we convolve the model spectrum with a Gaussian to match
the spectral resolution of GPI in K-band, and interpolate to the same
wavelengths of the GPI spectrum. We adjust the radius so that it minimizes
$\chi^2$ between the model and our spectra. The models are only matched to our
H and K spectra. We also show broadband photometry \citep{marois2008,
marois2010, galicher2011, currie2011, skemer2012, skemer2014, currie2014,
zurlo2016}, previously compiled in \cite{bonnefoy2016}, leaving out SPHERE
H-band points, which are slightly discrepant from our spectra. Table
\ref{tab:models} summarizes model parameters fit to each planet. Figure
\ref{fig:allmodels} displays each model presented in the table alongside our
spectrum and photometry from literature.  Each set of models is discussed in
detail in the following sections. 

\begin{table*}
\caption{Best fitting models \label{tab:models}}
\begin{center}
\begin{tabular}{|l|c|c|c|c|c|}
\hline
    Planet & Model & Radius ($M_{Jup}$) & T$_{eff}$ (K) & log(g) & $\log \sigma
T_{eff}^4 4\pi R^2 /L_{\odot}$ \\ \hline\hline
    HR 8799 c & PHOENIX (v16) & 1.2 & 1100 & 3.5 & -4.72 \\
             & Saumon+ (2008) \textit{fixed} & 1.4 & 1100 & 4.0 &  -4.58 \\
             & Saumon+ (2008) & 0.8 & 1300 & 3.75 & -4.78 \\
             & Cloud-AE60 & 0.75 & 1300 & 3.5 & -4.83 \\
             & BT-Settl & 0.7 & 1350 & 3.5 & -4.83 \\
    HR 8799 d & PHOENIX (v16) & 1.2 & 1100 & 3.5 & -4.72 \\
             & Saumon+ (2008) \textit{fixed} & 1.4 & 1100 & 4.0 & -4.58 \\
             & Saumon+ (2008) & 0.8 & 1300 & 4.0 & -4.78 \\
             & Cloud-AE60 & 0.65 & 1400 & 3.5 & -4.83 \\
             & BT-Settl & 0.65 & 1600 & 3.5 &  -4.60 \\
    HR 8799 e & PHOENIX (v16) & 1.3 & 1100 & 3.5 &  -4.65 \\
             & Saumon+ (2008) \textit{fixed} & 1.4 & 1100 & 4.0  & -4.58 \\
             & Saumon+ (2008) & 0.9 & 1300 & 3.75 &  -4.68 \\
             & Cloud-AE60 & 1.15 & 1100 & 3.5 &  -4.75 \\
             & BT-Settl & 0.6 & 1650 & 3.5  & -4.61 \\
             \hline
\end{tabular}
\end{center} \end{table*} 

\subsection{Published Best-Fit Models}

\subsubsection{PHOENIX model} \label{sec:phoenix}
The PHOENIX (v16) models from \cite{barman2011} are a set of parameterized
models with clouds, where clouds consist of a complex mixture or particles
whose state depend on temperature and pressure. This set of models takes into
account the transition between cloudless and cloudy atmosphere, as seen between
L- and T-type objects. The model is designed to identify the major physical
properties of the atmosphere. 

\cite{konopacky2013} presented a best fit model to a Keck/OSIRIS spectrum of HR
8799 c by fitting wavelength ranges and features that were most sensitive to
each model parameter (such as gravity, effective temperature, and cloud
thickness), checking consistency with broadband photometry. A combination of
dynamical stability, age, and interior structure models restricted the fit to
$\lesssim3.5 <\mathrm{log}~g \lesssim 4.4$ and $900K \lesssim T_{eff} \lesssim
1300 K$, leading to a model at $\mathrm{log}~g=4.0$ and $T_{eff}=1100K$,
moderate cloud thickness, a large eddy diffusion coefficient $K_{zz}=10^8
cm^2s^{-1}$, and super-solar C/O. 

We normalize (by scaling the radius) the model to fit each spectrum and show it
as the dashed line in Figure \ref{fig:allmodels}.  We note that this model was
fit to a higher resolution spectrum and still provides a good match to both our
lower resolution K-band spectrum of c as well as the new H-band spectrum. 
While the model was fit to the spectrum of HR 8799 c we show it alongside all
three spectra for comparison.  Based on our comparison in \S \ref{sec:chi2}
(Figure \ref{fig:hists}) we do not expect it to provide a good fit for d, but
could possibly fit e within errorbars. We see that this model does not capture
the shape of the K-band spectrum of d nor the flatter H-band spectrum. The
model is reasonably consistent with e, within the large errobars.  This model,
while scaled just to our H \& K spectra, is most consistent with the $3-5\mu m$
photometry in all 3 cases. This highlights the importance of obtaining spectra,
which will show more detailed differences between objects and can better
distinguish between models.

\subsubsection{Patchy cloud model \citep{saumonmarley2008}} \label{sm}
Models from \cite{saumonmarley2008} are evolution models for brown dwarfs and
giant planets in the ``hot start" scenario that include patchy clouds. These
are parameterized based on effective temperature, cloud properties (cloud hole
fraction), gravity, and mixing properties, namely a sedimentation parameter
$f_{sed}$ defined in \cite{ackerman2001}, the ratio between the sedimentation
velocity and the convective velocity scale. 
\citet{ingraham2014} fit Patchy Cloud models to GPI spectra of c and d, in two
cases:  first with fixed radius (based on evolutionary models), and then with
the radius allowed to vary. 
The fixed-radius models both had $T_{eff}=1100K$ with thick clouds including
some horizontal variation to account for observed J-band flux. The model that
provided the best fit for c has $f_{sed}$ of 0.25 and cloud hole fraction of
5\% and the best fit for d has $f_{sed}$=0.50 and no holes. The free-radius
models both have $T_{eff}=1300K$, but require a radius of $<1 R_{Jup}$.  The
best fit for c has $f_{sed}$=1, without cloud holes and the best fit for d has
$f_{sed}=0.5$ and 5\% holes.

\cite{ingraham2014} found that while fixed-radius models are able to reproduce
the planet broadband SEDs, they were consistent with the more detailed spectra.
We show the described models alongside the spectra of all three planets (Figure
\ref{fig:allmodels}). The best fitting models for c provided a better fit to
HR8799 e and we show this alongside the e spectrum.  Similar to
\citet{ingraham2014}, we find that the free-radius model provides a better fit
to both the H and K band spectra, but not the $3-5\mu m$ photometry, and that
while the fixed-radius model fits broadband photometry, it does not provide a
good fit to our spectra. For c and d fitting both the H and K spectrum for the
best scaling leads to a poorer fit of the K-band portion, and this is
especially obvious for the longer wavelength part of d's K band spectrum. For
e, while both models match the peak flux at H and K within the large error
bars, they do not capture the band edges. In general, the models do not capture
the relative flux between H and K in all cases.

The free-radius models may be missing some effect that leads to requiring
sub-$R_{Jup}$ radii to match the observed spectrum. Modeling these objects may
require one or a combination of clouds, non-equilibrium chemistry, and
non-solar metallicity to be consistent with both the broadband SED and
spectroscopy. 

\subsubsection{A Note on Composition}

\citet{konopacky2013} found margnial evidence for higher C/O ratio for HR8799 c
compared to the host star, which has implications on the planet formation
history as suggested by \citet{oberg2011}. Without detailed high resolution
spectra our results cannot constrain abundances, but it is encouraging that the
same model, based on the strongest spectral line indicators also provides a
good fit for our lower resolution spectrum and new H-band data. Our data do
show evidence of clear differences between the spectra, suggesting there could
be differences between the compositions. These differences could also be the
result of first order physical effects, rather than composition.

\citet{lavie} attempted to recover abundances of the HR8799 planets through
atmospheric retrieval. Given the differences we see in our spectra, including
new data, some of the differences seen in \cite{lavie} could be real. We
additionally present new K-band data for e, which they conclude is require to
estimate C/O and C/H ratios. However the issue of unrealistic radii, as
discussed in their study remains to be an unsolved problem of modeling. Until
atmospheric models can provide a physically motivated reason for lower observed
flux (that often leads to reducing the radius). More spectroscopy, especially
at higher resolution, could help identify the dominant effect, whether clouds
\citep[e.g.,][]{saumonmarley2008, burningham2017}, non-equilibrium chemistry
\cite[e.g.,][]{barman2011}, composition \citep[e.g.,][]{lee2013}, atmospheric
processes \citep[e.g.,][]{tremblin2017}, or some combination. JWST near-IR and
mid-IR spectroscopy could help resolve what physical mechanism drives this
effect to more accurately determine atmospheric compositions.

\subsection{Model grids}

\subsubsection{CloudAE-60 model grid} \label{sec:cloudae}
We consider the CloudAE-60  model grid \citep{madhusudhan2011}, also discussed
in \cite{bonnefoy2016}. These models represent thick forsterite clouds at solar
metallicity with mean particle size of 60$\mu m$. These models do not account
for disequilibrium chemistry. We fit the grid of models between 1100 and 1600K
and scale to the best fitting radius. In Figure \ref{fig:allmodels} we plot the
CloudAE-60 model that minimizes $\chi^2$.

This set of models is able to reproduce the K-band spectra of c and e fairly
well all the way to band edges. The model does not match the shape of the d
spectrum, neither representing the flatter H-band spectrum, nor the rising
K-band spectrum.  We find similar best fitting effective temperature for e as
in \cite{bonnefoy2016}.  All three cases produce models that require radii
below $1 R_{Jup}$. The models that best-fit the H \& K spectra do not match the
flux at 3-5 $\mu$m.

\subsubsection{BT-Settl model grid} \label{sec:btsettl}
Lastly, the BT-Settl 2014 evolutionary model grid for low mass stars couples
atmosphere and interior structures \citep{baraffe2015}. We consider a
temperature range from $1200 - 1700$K and gravity range $\log{g} =3.0 - 4.0$,
encompassing the best fits shown in \cite{bonnefoy2016}. The grid provides
models in steps of 100K; to estimate intermediate temperatures, we average
models to search in steps 50K. We show these best fit parameters in Figure
\ref{fig:allmodels}.

We find similar best fitting effective temperatures and gravity as
\cite{bonnefoy2016}. This model better reflects the rising slope in K for
planet d and in this case is roughly consistent with photometry beyond 3$\mu$m.
These models also under-predict flux from 3-5$\mu$m in some cases.
\cite{bonnefoy2016} similarly noted that this model did not match both the Y-H
spectra and the 3-5 $\mu$m flux, possibly indicating that it does not produce
enough dust at high altitudes. In both studies this model matches the planet d
photometry better than for c.

\section{Summary and Conclusions} \label{sec:conclusions}

We have implemented a forward modeling approach to recovering IFS spectra from
GPI observations of the HR 8799 planets c, d and e. Using this approach we have
re-reduced data first presented in \cite{ingraham2014} as well as new H-band
data with this new algorithm, finding as in \cite{pueyo2016} that algorithm
parameters converge with increasing $k_{klip}$. With this approach we are able
to recover a K-band spectrum on HR 8799 e for the first time.  While the HR
8799 planet SEDs have been typically shown to be very similar, their more
detailed spectra show evidence of different atmospheric properties. In addition
to showing that there is statistical difference between c and d, different
atmospheric models also provide best fits to each spectrum. These differences
could be the result of properties such as cloud fraction, non-equilibrium
chemistry, composition, and/or thermal structure. We have shown that a range of
models with difference physical mechanisms can provide similar fits to our H \&
K spectra. 

While the large errorbars of HR 8799 e make it hard to determine its similarity
to the other two planet spectra, but we find that c and d are distinct.
The dominant effect comes from the relative flux between H and
K bands; the differences go away when we normalize the spectra in each band
individually (see Appendix \ref{sec:chi2_by_band}). With less noisy K-band
data we could make a stronger statement about difference between these and
planet e. It is likely the K-band spectrum for e resembles that of d and that
our K1 band edge is biased by residual speckle noise. Given the mid-late
L-spectral types of all three planets, they may display variability.  Planet e
could be a good candidate for variability study, since it appears to have a
brighter H-band spectrum, suggestive of cloud holes in transition from L-T
\citep[e.g.,][]{radigan2014}. 

The following summarize the main points of our results:
\begin{itemize}
\item The KLIP-FM method is able to recover a spectrum of close-in planet e,
despite many residual speckles. By exploring results over varying $k_{klip}$
and $mov$ we the relatively low dependence of our result on algorithm
parameters. This is especially true for the H-band data, which had greater
field rotation. 
\item Our H spectra of planets d and e are consistent at the short end with YJH
spectra from the SPHERE/VLT instrument \citep{zurlo2016}.
\item The H \& K spectrum of HR 8799 c is statistically different ($>5\sigma$)
from d based on our $\chi^2_{i,j}$ measurement. 
\item All three objects are best matched by mid to late L-type field brown
dwarfs from a library of near-infrared spectra. Evidence of L-T transition
variability could support models with inhomogeneous cloud coverage.
\item The PHOENIX model, which was fit to higher resolution K-band spectroscopy
of HR 8799 c also fits both our lower resolution H and K-band spectra, as well
as 3-5 $\mu$m, without requiring sub-$M_{Jup}$ radii. We have compared it to
the spectra of d and e as a reminder that even if the broadband photometry is
similar, the same model will not provide a good fit for all objects in the
system. 
\item A general grid of models is not expected to provide the same level of
detailed fit as a detailed study, however the CloudAE models produce very
similar results for our H-K spectra as the PHOENIX model, while invoking
different mechanisms. The BT-Settl models seem to represent the H \& K spectra
of d best, while also matching $3-5 \mu m$ photometry. The model grids also
require sub-$M_{Jup}$ radii in most cases.
\end{itemize}

Spectroscopic information is necessary for revealing compositional differences
between the HR 8799 planets. However, more work is needed to accurately measure
abundances.  
That most models require unreasonable radii indicate that they are missing a
physical mechanism to fully account for the observed flux, similarly discussed
in many previous studies as an ``under-luminosity problem" that indicates a
discrepancy between  atmospheric models and evolutionary models
\citep[e.g.,][]{marois2008, bowler2010, marley2012}.  While this problem could
be accounted for by a prescription of clouds \citep[e.g.,][]{madhusudhan2011},
chemistry \citep[e.g.,][]{barman2011}, or thermal structure
\citep[e.g.,][]{tremblin2017}, the true process is not solved.  When the
correct physical mechanism is understood, modeling and/or atmosphere retrievals
should be able to measure abundances accurately.  Larger spectral coverage
could help distinguish between different processes.  Higher spectroscopic
resolution would provide more detail to probe atmospheric chemistry, which
could be achieved with next generation Extremely Large Telescopes (ELTs) that
combine high resolution spectroscopy with high contrast imaging \citep[as
described in][]{snellen2015}.

JWST will be able to deliver spectra at longer wavelengths for better
characterization planetary mass companions at 3-5$\mu$m and beyond. Given the
reduced inner working angle and limited rotation compared to ground-based
8m-class telescopes, forward modeling may be important for obtaining infrared
companion spectra to systems like HR 8799, helping to advance atmospheric
modeling efforts and provide benchmark objects for future high contrast imaging
studies. 

\acknowledgments

We thank the anonymous referee for helpful comments. This work is based on
observations obtained at the Gemini Observatory, which is operated by the
Association of Universities for Research in Astronomy, Inc., under a
cooperative agreement with the NSF on behalf of the Gemini partnership: the
National Science Foundation (United States), the National Research Council
(Canada), CONICYT (Chile), Ministerio de Ciencia, Tecnolog\'{i}a e
Innovaci\'{o}n Productiva (Argentina), and Minist\'{e}rio da Ci\^{e}ncia,
Tecnologia e Inova\c{c}\~{a}o (Brazil).
This research has made use of the SVO Filter Profile Service
(http://svo2.cab.inta-csic.es/theory/fps/) supported from the Spanish MINECO
through grant AyA2014-55216. 
This research has benefited from the SpeX Prism Library maintained by Adam
Burgasser at http://www.browndwarfs.org/spexprism. This research has made use
of NASA's Astrophysics Data System and the AstroBetter blog and wiki.
Work from A.Z.G was supported in part by the National Science Foundation
Graduate Research Fellowship Program under Grant No. DGE1232825. A.Z.G and A.S.
acknowledge support from NASA grant APRA08-0117 and the STScI Director’s
Discretionary Research Fund.
The research was supported by NSF grant AST-1411868 and NASA grant NNX14AJ80G
(J.-B.R.).
P.K., J.R.G., R.J.D., and J.W. thank support from NSF AST-1518332, NASA
NNX15AC89G and NNX15AD95G/NEXSS.  This work benefited from NASA’s Nexus for
Exoplanet System Science (NExSS) research coordination network sponsored by
NASA's Science Mission Directorate. A.B. acknowledges support for this research
under their NASA WFIRST-SIT award \# NNG16PJ24C, NASA Grant NNX15AE19G, and
NASA JPL subcontracts no. 1538907, 1529729, 1513640 \& 1534432-B-4.25.  

\facility{Gemini-S}.

\software{PyKLIP \citep{pyklip}, 
          Astropy \citep{2013A&A...558A..33A},  
          Matplotlib \citep{matplotlib},
          Numpy \& Scipy \citep{numpy}
          }

\appendix

\section{Residuals} \label{sec:residuals}
\begin{figure}[htbp]
    \centering
    \includegraphics[scale=0.33]{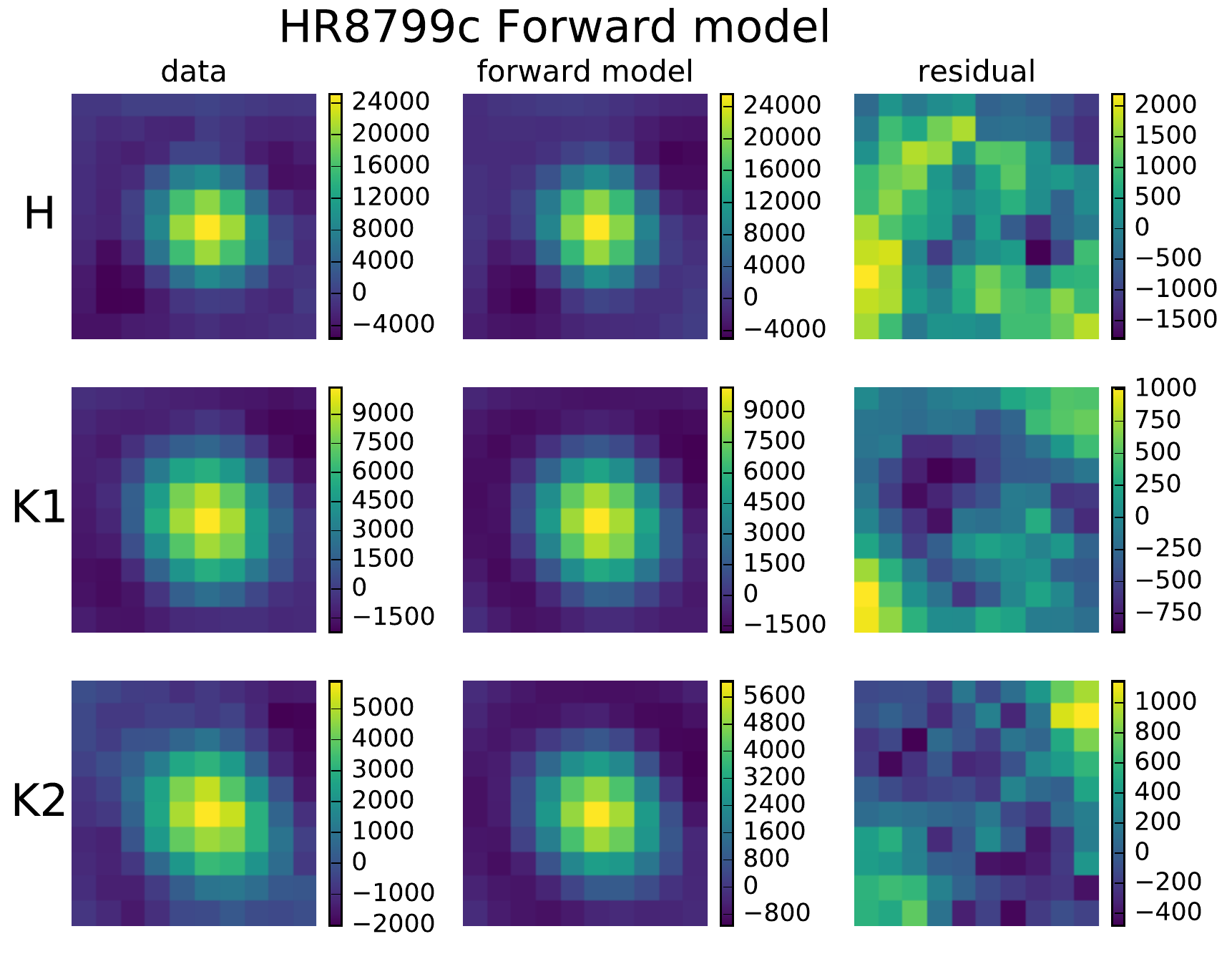}
    \includegraphics[scale=0.33]{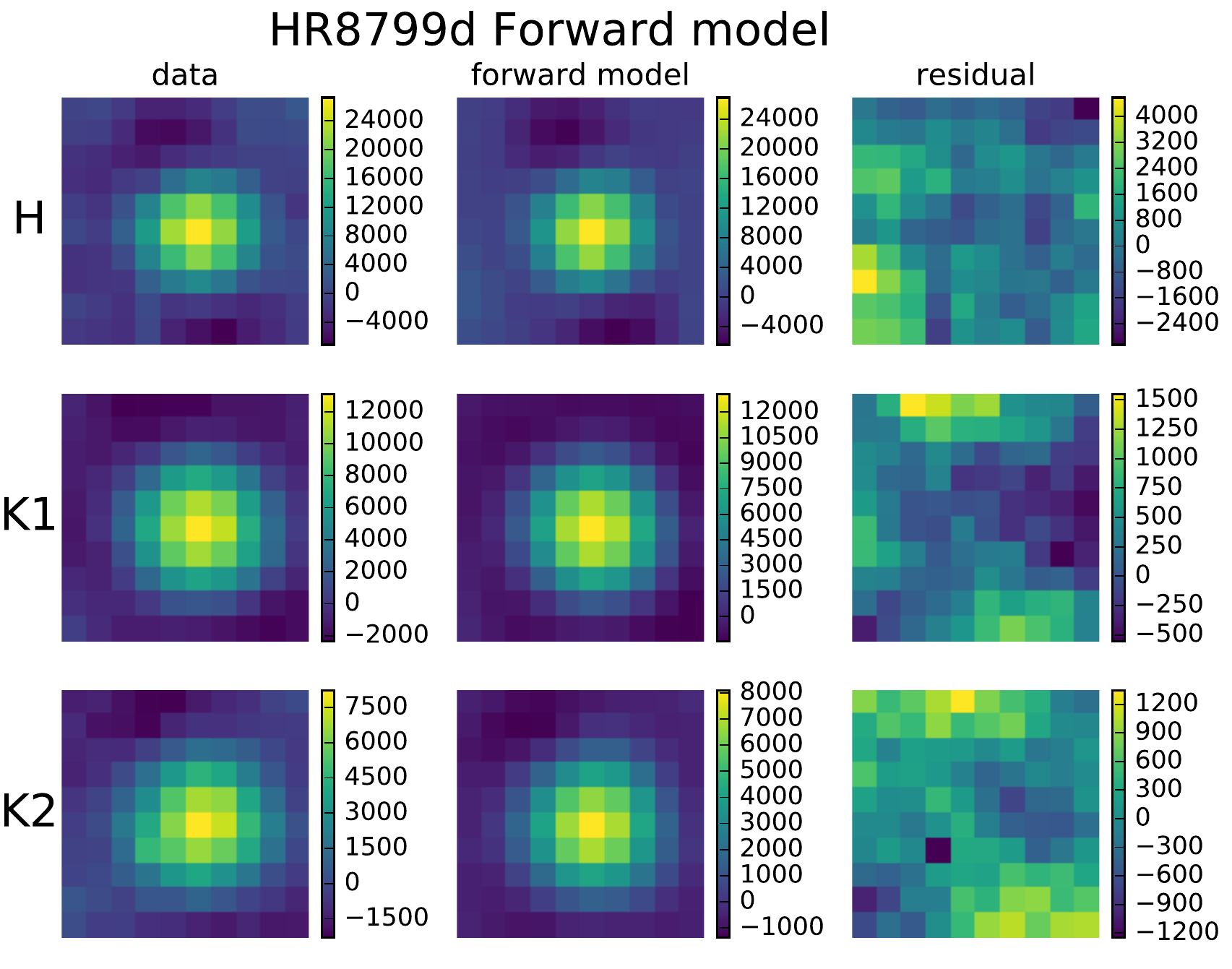}
    \includegraphics[scale=0.33]{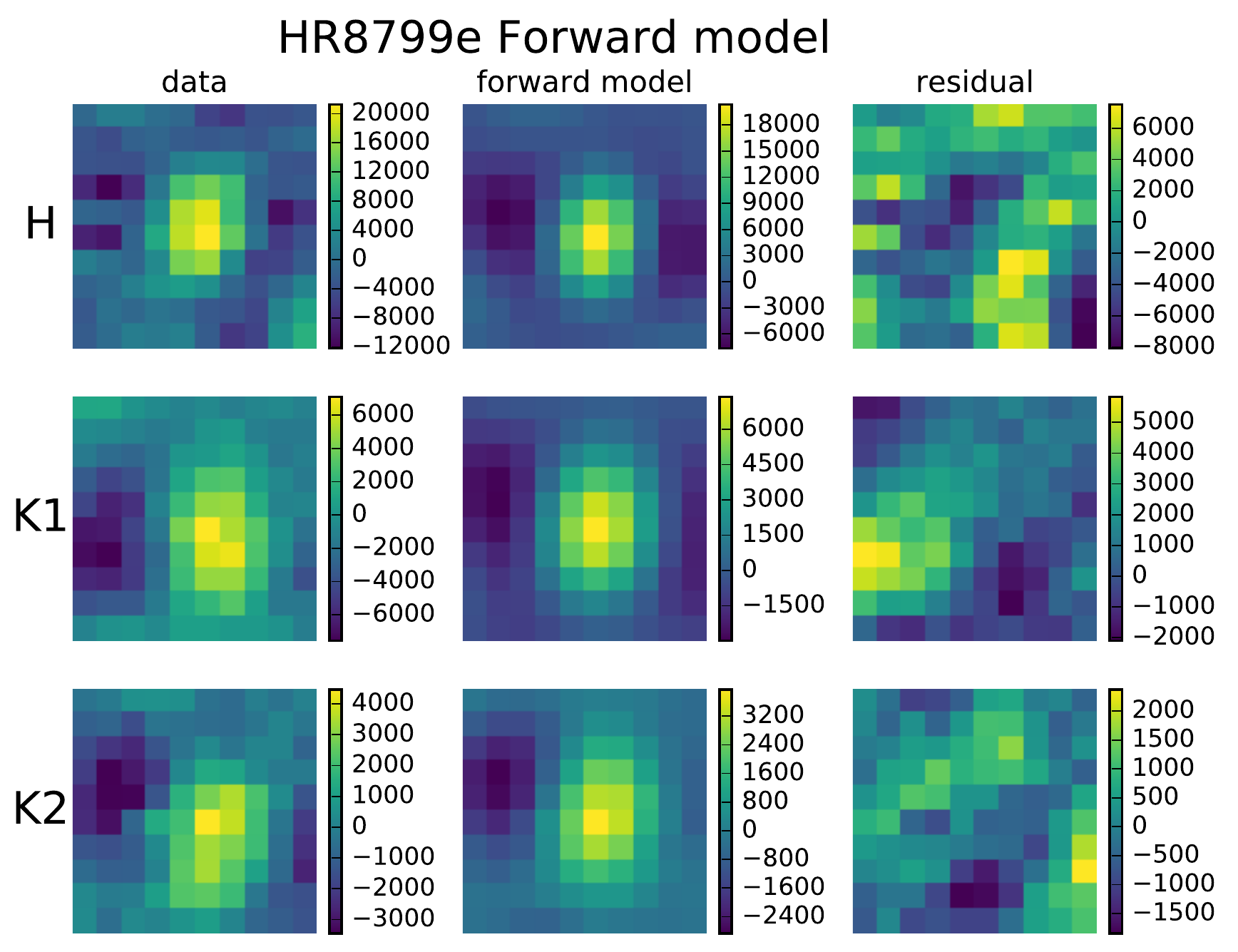}
    \caption{A summary of the forward model performance for HR 8799 c (left) d
(middle) and e (right). Klipped data is shown to the left, the forward model is
in the middle panel and the rightmost panel shows the difference between these
two.  Each image stamp is summed over the bandpass. Pixel values are in raw
data units. Residual speckle noise appears to be influencing the K1 reduction
of HR 8799 e.}
    \label{fig:residuals}
\end{figure}

After optimizing KLIP parameters, described in section \ref{sec:klip}, we
display the PSF-subtracted data stamp that contains the planet and the forward
model, summed over the bandpass, and the residual between the two (Figure
\ref{fig:residuals}). We show the residual plots for all three planets in each
H, K1, and K2 bands, summed over the wavelength axis. 

Increased speckle noise close to the focal plane mask of the image contributes
larger residuals for the forward model of HR 8799 e. However, as can be seen in
the Figure \ref{fig:residuals}, the forward model captures the self-subtraction
negative lobes. We have taken care to remove obvious contaminants, including a
bad pixel in several wavelength channels near HR 8799 d.  

\section{Comparing algorithms} \label{sec:algos}

We compare our spectral extraction with the reductions from two other
algorithms. We expect KLIP-FM to perform well especially at small inner working
angle and when the companion is faint. The recovery of the more widely
separated c and d planets provide a good validation of the forward modeling
algorithm. Some flux loss is expected when the forward model assumptions are
not completely appropriate, as described in \cite{pueyo2016} and as we have
discussed in \S \ref{sec:klip}.

\begin{figure}[htbp!]
    \centering
    \includegraphics[scale=0.6]{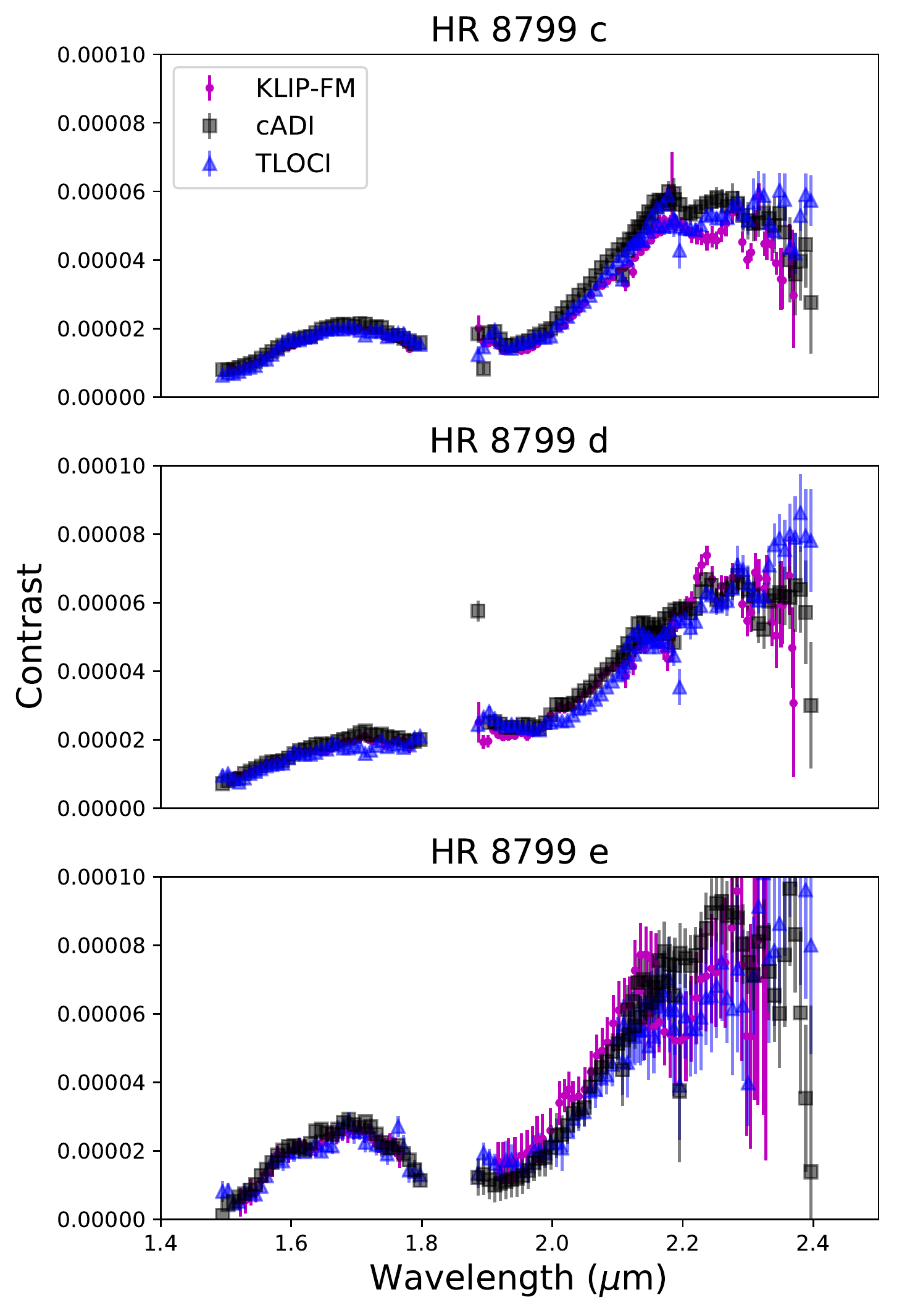}
    \caption{Comparison of the KLIP forward modeling algorithm reduction
presented in this paper (magenta circles) with cADI (black squares) and TLOCI
(blue triangles) reductions. The planet-to-star contrast is plotted vs.
wavelength.}
    \label{fig:algos}
\end{figure}

The two PSF subtraction pipelines we compare with were both used for the
discovery of 51 Eri b and are fully described in \citet{macintosh2015}. These
use the cADI and TLOCI algorithms, respectively.  We show results from the
three pipelines together in Figure \ref{fig:algos}. We briefly recall the
important steps for each pipeline, which starts from the same calibrated
datacubes generated by the Data Cruncher \citep{wang2018}, and perform the PSF
subtraction and signal extraction.

In the first pipeline the images are high-pass filtered with an apodized
Fourier-space filter following a Hanning profile. The filtering is done early
in the process to simultaneously affect both satellite spots and planets.  We
tested different cutoff frequencies were tested. We found planet c to be
particularly sensitive the size of the filter in the K band datasets.  Being
brighter than the spot in a speckle-free region of the image, its absolute flux
was hard to calibrate based on the residuals after forward-modeling through
inserting simulated signals. The best results were found with a cutoff
frequency of 8 equivalent-pixels in the image plane for planets c and d. For
planet e which is heavily embedded in the speckle field, a more agressive
filter with a cutoff frequency of 4 equivalent-pixels was applied. At each
wavelength, a model of the PSF for the forward model was built from the average
of the four spots in each image, further combined along the time dimension. The
PSF subtraction was applied with the cADI algorithm \citep{marois2006,
lagrange2010}. To accurately measure the position of each planet, the residuals
were stacked along the spectral dimension. The position was extracted with an
amoeba algorithm that minimizes the squared residuals, after subtracting the
forward-model to the planet'signal, in a $2\times2$ FWHM wedge centered on the
planet. The spectrum of each planet was then extracted with the same technique
but fixing the position of the forward model at the best fitted position.
Uncertainties were estimated from the dispersion of the recovered signal of
fake sources that were injected at the same separation but twenty different
position angles uniformly distribution between $90$ and $270^\circ$ apart from
each planet.

The second pipeline uses the TLOCI algorithm, which combines both SDI and ADI
data into one stellar subtraction step.  Generic methane/dusty input spectra
were used to guide the reference image selection process to minimize
self-subtraction and maximize the signal-to-noise for any companion with a
spectrum similar to the input spectra. A maximum flux contamination ratio of
90\% inside a 1.5 $\lambda/D$ aperture was chosen for this analysis.  The
algorithm uses a pixel mask to avoid fitting the planet flux with the algorithm
and an 11x11 pixel (3x3 $\lambda/D$) median high-pass filter.  Circular annulus
subtraction regions have 1.7 $\lambda/D$ width. The least-squares optimization
regions are also circular annuli just inside and outside the subtraction
regions, having 3 $\lambda/D$ width for the inner annulus and 6.6 $\lambda/D$
width for the outer annulus. To determine the companion spectrum, a
polychromatic forward model is generated from the least-squares coefficients
and the stellar PSF (obtained from median averaging the four off-axis
calibration spots) mimicking the exoplanet signal, including the negative
“wings” from self-subtraction. This model is adjusted in position (sub-pixel
accuracy) and flux using an iterative technique that minimizes the residual
local noise in an aperture after subtracting the forward model from the
exoplanet candidate signal. Photometric error bars are estimated at each
wavelength by taking the standard deviation of the same extraction process
performed on simulated exoplanets (using the same polychromatic forward model)
located at the exoplanet candidate separation, but at different position
angles.

Overall, the KLIP-FM spectra match the cADI and/or TLOCI spectra in most cases
as shown in Figure \ref{fig:algos}. The KLIP forward modeling algorithm and the
cADI-based pipeline show excellent agreement for planet d, for planet c at H
band, and at the $\lesssim1\sigma$ level for planet e. The cADI pipeline is
$1\sigma$ brighter at K band. This discrepancy is likely due to the impact of
different high-pass filters as discussed before.  The TLOCI and KLIP-FM
pipelines show the same excellent agreement for planet c and is slightly
fainter for planet d towards the end of H band. The TLOCI analysis agrees with
KLIP-FM within the large errorbars for planet e.

\section{Spectrum comparison by band} \label{sec:chi2_by_band}
\label{sec:extrahistograms}
\begin{figure}[htbp]
    \centering
    \includegraphics[scale=0.5]{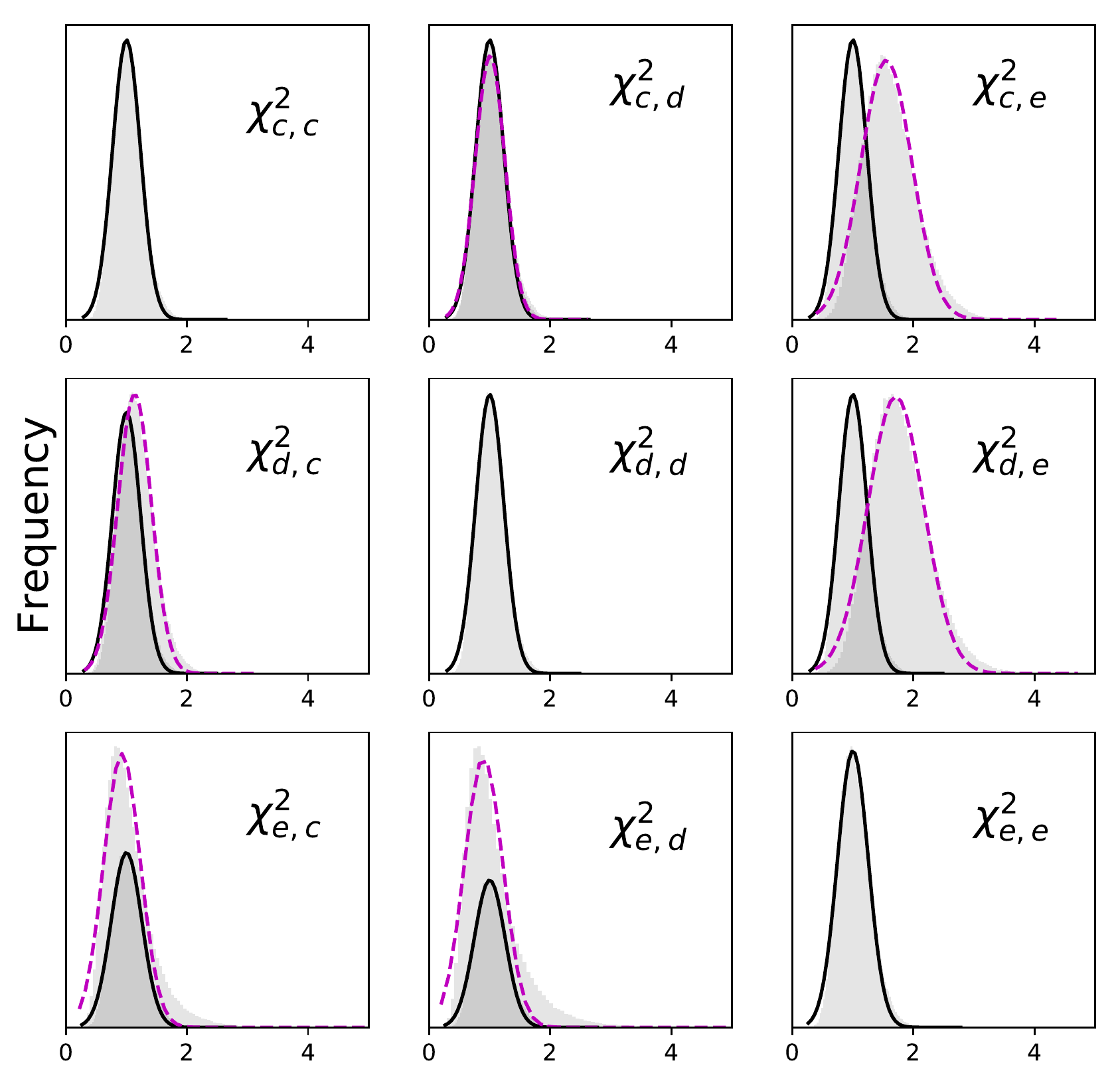}
    \caption{Comparison of $\chi^{2}$ distributions for H band.}
    \label{fig:indiv_hists_h}
\end{figure}

\begin{figure}[htbp]
    \centering
    \includegraphics[scale=0.5]{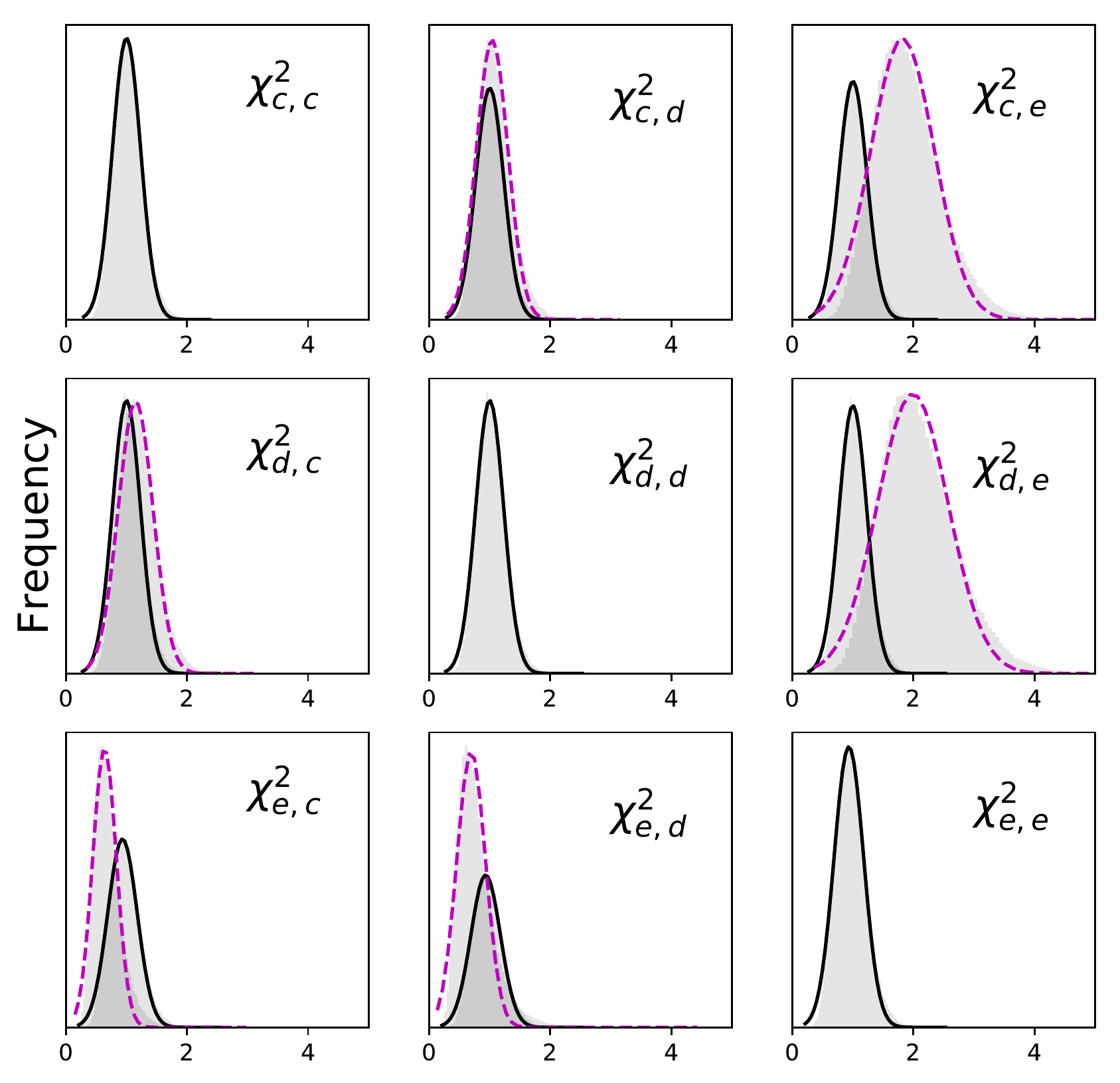}
    \caption{Comparison of $\chi^{2}$ distributions for K1 band.}
    \label{fig:indiv_hists_k1}
\end{figure}

\begin{figure}[htbp]
    \centering
    \includegraphics[scale=0.5]{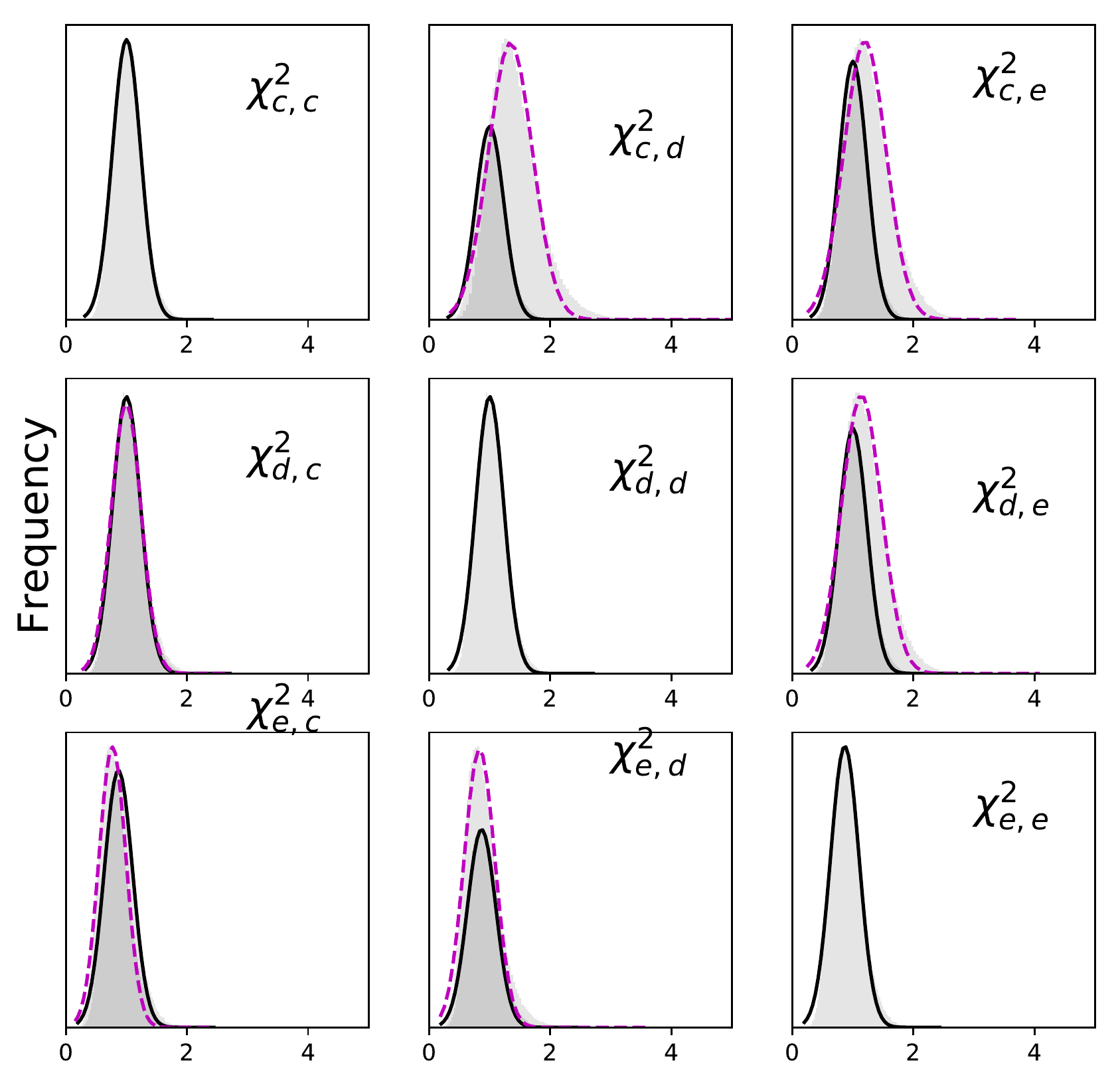}
    \caption{Comparison of $\chi^{2}$ distributions for K2 band.}
    \label{fig:indiv_hists_k2}
\end{figure}

We show the individual spectra comparisons between HR 8799 c, d, and e  as
described in Equation \ref{eqn:chi2hist}. Here we normalize the spectrum in
each band, rather than over the entire H-K range.  Figures
\ref{fig:indiv_hists_h}-\ref{fig:indiv_hists_k2} show the individual comparison
for each band.

The overall shapes of the spectra within each band are not very different
within errorbars. Figures \ref{fig:indiv_hists_h}-\ref{fig:indiv_hists_k2} show
considerable overlap in the distributions. However, the relative flux between H
and K bands for the three planets is an obvious difference between them (which
also drives the atmospheric model fitting) that is not captured by the
individual band comparison of low resolution spectra. 

\section{Full reduced spectrum} \label{sec:spectrum}
We provide our $1.5-2.4\mu m$ reduced spectrum in Table \ref{tab:spectrum}.

\startlongtable
\begin{deluxetable*}{lcclcclcc} 
\tablecaption{GPI spectrum of the HR 8799 c, d, and e in flux units at 10pc. \label{tab:spectrum}}
\tablehead{ \colhead{\textbf{c:} $\lambda$} & \colhead{Flux } & 
\colhead{Error} & \colhead{\textbf{d:}  $\lambda$ } &
 \colhead{Flux} & \colhead{Error} & \colhead{\textbf{e:}  $\lambda$} 
& \colhead{Flux} & \colhead{Error} \\
\colhead{($\mu m$)} & \colhead{($W m^{-2} \mu m^{-1}$)} 
& \colhead{($W m^{-2} \mu m^{-1}$)} &
\colhead{($\mu m$)} & \colhead{($W m^{-2} \mu m^{-1}$)}  & \colhead{($W m^{-2} \mu m^{-1}$)}  &
\colhead{($\mu m$)}  & \colhead{($W m^{-2} \mu m^{-1}$)}  & \colhead{($W m^{-2} \mu m^{-1}$)} }
\startdata
1.50776 & 1.73145E-15 & 2.79261E-16 & 1.50776 & 1.33467E-15 & 2.88152E-16 & 1.50776 &  &   \\ 
1.51072 & 1.46075E-15 & 2.55719E-16 & 1.51072 & 1.44751E-15 & 2.50840E-16 & 1.51072 &  &   \\ 
1.51495 & 1.56397E-15 & 2.62620E-16 & 1.51495 & 1.88622E-15 & 2.62037E-16 & 1.51495 &  &   \\ 
1.52168 & 1.59637E-15 & 2.71240E-16 & 1.52168 & 1.85274E-15 & 2.54648E-16 & 1.52168 & 1.00613E-15 & 8.32404E-16  \\ 
1.52962 & 1.82500E-15 & 2.63159E-16 & 1.52962 & 1.88055E-15 & 2.59543E-16 & 1.52962 & 1.10141E-15 & 7.53412E-16  \\ 
1.53807 & 1.89682E-15 & 2.51584E-16 & 1.53807 & 1.98775E-15 & 2.54490E-16 & 1.53807 & 1.62327E-15 & 7.52152E-16  \\ 
1.54601 & 1.92735E-15 & 2.43479E-16 & 1.54601 & 2.09302E-15 & 2.44374E-16 & 1.54601 & 1.74446E-15 & 5.72797E-16  \\ 
1.55387 & 1.95888E-15 & 2.23333E-16 & 1.55387 & 2.13862E-15 & 2.31782E-16 & 1.55387 & 2.44538E-15 & 4.77751E-16  \\ 
1.56234 & 2.20817E-15 & 2.26991E-16 & 1.56234 & 2.30514E-15 & 2.45245E-16 & 1.56234 & 2.64150E-15 & 5.18419E-16  \\ 
1.57074 & 2.28764E-15 & 2.15666E-16 & 1.57074 & 2.33754E-15 & 2.43456E-16 & 1.57074 & 2.75700E-15 & 5.71800E-16  \\ 
1.57898 & 2.38194E-15 & 2.07432E-16 & 1.57898 & 2.36517E-15 & 2.49826E-16 & 1.57898 & 3.37050E-15 & 4.86809E-16  \\ 
1.58728 & 2.39028E-15 & 1.98210E-16 & 1.58728 & 2.22597E-15 & 2.46330E-16 & 1.58728 & 3.20412E-15 & 3.65439E-16  \\ 
1.5957  & 2.50643E-15 & 2.04841E-16 & 1.5957  & 2.49010E-15 & 2.51572E-16 & 1.5957  & 3.23819E-15 & 4.05298E-16  \\ 
1.60374 & 2.51859E-15 & 1.83617E-16 & 1.60374 & 2.49566E-15 & 2.07153E-16 & 1.60374 & 3.50249E-15 & 4.09458E-16  \\ 
1.61209 & 2.71523E-15 & 1.96294E-16 & 1.61209 & 2.68137E-15 & 2.17858E-16 & 1.61209 & 3.64782E-15 & 3.79600E-16  \\ 
1.62078 & 2.73097E-15 & 1.89629E-16 & 1.62078 & 2.71290E-15 & 1.90383E-16 & 1.62078 & 3.53528E-15 & 3.34049E-16  \\ 
1.62895 & 2.69177E-15 & 1.70791E-16 & 1.62895 & 2.64639E-15 & 2.09287E-16 & 1.62895 & 3.24040E-15 & 3.56117E-16  \\ 
1.63718 & 2.68906E-15 & 1.63056E-16 & 1.63718 & 2.56599E-15 & 2.12826E-16 & 1.63718 & 3.11055E-15 & 3.63156E-16  \\ 
1.646   & 2.81385E-15 & 1.64888E-16 & 1.646   & 2.40662E-15 & 2.29716E-16 & 1.646   & 3.31398E-15 & 3.81471E-16  \\ 
1.65415 & 2.85857E-15 & 1.68992E-16 & 1.65415 & 2.52209E-15 & 2.42221E-16 & 1.65415 & 3.69580E-15 & 4.54456E-16  \\ 
1.66191 & 2.91661E-15 & 1.68455E-16 & 1.66191 & 2.62517E-15 & 2.04352E-16 & 1.66191 & 3.61348E-15 & 4.12620E-16  \\ 
1.6697  & 2.84368E-15 & 1.57134E-16 & 1.6697  & 2.62696E-15 & 1.72916E-16 & 1.6697  & 3.52549E-15 & 4.24976E-16  \\ 
1.67826 & 2.69356E-15 & 1.50773E-16 & 1.67826 & 2.61324E-15 & 1.59223E-16 & 1.67826 & 3.45127E-15 & 3.69226E-16  \\ 
1.68659 & 2.53747E-15 & 1.58609E-16 & 1.68659 & 2.62474E-15 & 1.52072E-16 & 1.68659 & 3.29716E-15 & 3.95526E-16  \\ 
1.69548 & 2.67212E-15 & 1.58928E-16 & 1.69548 & 2.74329E-15 & 1.60122E-16 & 1.69548 & 3.59782E-15 & 3.86855E-16  \\ 
1.70379 & 2.79548E-15 & 1.75119E-16 & 1.70379 & 2.81561E-15 & 1.83871E-16 & 1.70379 & 3.62358E-15 & 3.96528E-16  \\ 
1.71256 & 2.68000E-15 & 1.91466E-16 & 1.71256 & 2.84797E-15 & 1.91948E-16 & 1.71256 & 3.38168E-15 & 4.28278E-16  \\ 
1.71826 & 2.63503E-15 & 1.72197E-16 & 1.71826 & 2.61451E-15 & 1.78198E-16 & 1.71826 & 3.30321E-15 & 3.66134E-16  \\ 
1.72562 & 2.49994E-15 & 1.58232E-16 & 1.72562 & 2.47666E-15 & 1.46859E-16 & 1.72562 & 2.80979E-15 & 3.23274E-16  \\ 
1.73395 & 2.27136E-15 & 1.38421E-16 & 1.73395 & 2.34448E-15 & 1.37585E-16 & 1.73395 & 2.34737E-15 & 3.33969E-16  \\ 
1.74305 & 2.16220E-15 & 1.25966E-16 & 1.74305 & 2.19943E-15 & 1.25607E-16 & 1.74305 & 2.43047E-15 & 3.08768E-16  \\ 
1.75139 & 2.18413E-15 & 1.55635E-16 & 1.75139 & 2.32768E-15 & 1.39663E-16 & 1.75139 & 2.64743E-15 & 3.34755E-16  \\ 
1.75923 & 2.09152E-15 & 1.57566E-16 & 1.75923 & 2.48106E-15 & 1.42337E-16 & 1.75923 & 2.37151E-15 & 3.60365E-16  \\ 
1.76602 & 2.03956E-15 & 1.51567E-16 & 1.76602 & 2.38054E-15 & 1.52851E-16 & 1.76602 & 1.94834E-15 & 3.88357E-16  \\ 
1.77172 & 1.93018E-15 & 1.43464E-16 & 1.77172 & 2.33128E-15 & 1.92200E-16 & 1.77172 &  &   \\ 
1.77686 & 1.85298E-15 & 1.41944E-16 & 1.77686 & 2.21941E-15 & 2.01857E-16 & 1.77686 &  &   \\ 
1.78121 & 1.63147E-15 & 1.37720E-16 & 1.78121 & 2.07282E-15 & 2.04131E-16 & 1.78121 &  &   \\ 
1.88723 & 2.03801E-15 & 1.36498E-15 & 1.88723 & 2.30577E-15 & 1.44160E-15 & 1.88723 &  &   \\ 
1.89436 & 1.47770E-15 & 2.71668E-16 & 1.89436 & 1.88868E-15 & 2.94845E-16 & 1.89436 &  &   \\ 
1.90174 & 1.47124E-15 & 1.43759E-16 & 1.90174 & 1.88607E-15 & 2.12447E-16 & 1.90174 &  &   \\ 
1.9096  & 1.54153E-15 & 1.45717E-16 & 1.9096  & 2.09433E-15 & 1.66364E-16 & 1.9096  &  &   \\ 
1.91714 & 1.34378E-15 & 1.39897E-16 & 1.91714 & 1.99324E-15 & 1.44480E-16 & 1.91714 & 2.31105E-15 & 5.16806E-16  \\ 
1.92461 & 1.25976E-15 & 1.27087E-16 & 1.92461 & 1.99658E-15 & 1.48204E-16 & 1.92461 & 2.09120E-15 & 5.32716E-16  \\ 
1.93272 & 1.18958E-15 & 1.16669E-16 & 1.93272 & 1.92186E-15 & 1.37709E-16 & 1.93272 & 1.99156E-15 & 5.68846E-16  \\ 
1.94265 & 1.12703E-15 & 9.82724E-17 & 1.94265 & 1.76872E-15 & 1.17088E-16 & 1.94265 & 1.89713E-15 & 5.07610E-16  \\ 
1.95281 & 1.13187E-15 & 1.13328E-16 & 1.95281 & 1.83061E-15 & 1.36893E-16 & 1.95281 & 1.96871E-15 & 5.13199E-16  \\ 
1.96161 & 1.17873E-15 & 1.13429E-16 & 1.96161 & 1.86198E-15 & 1.39680E-16 & 1.96161 & 1.96727E-15 & 5.34209E-16  \\ 
1.96938 & 1.28330E-15 & 1.08946E-16 & 1.96938 & 1.84856E-15 & 1.22501E-16 & 1.96938 & 1.96955E-15 & 5.17002E-16  \\ 
1.97667 & 1.32508E-15 & 1.04380E-16 & 1.97667 & 1.78353E-15 & 1.10631E-16 & 1.97667 & 2.11890E-15 & 5.43009E-16  \\ 
1.98473 & 1.41538E-15 & 1.07500E-16 & 1.98473 & 1.88182E-15 & 1.28567E-16 & 1.98473 & 2.18175E-15 & 5.51286E-16  \\ 
1.99717 & 1.44730E-15 & 1.07762E-16 & 1.99717 & 2.06136E-15 & 1.54830E-16 & 1.99717 & 2.30878E-15 & 5.05001E-16  \\ 
2.0109  & 1.57343E-15 & 9.40893E-17 & 2.0109  & 2.17861E-15 & 1.47850E-16 & 2.0109  & 3.04514E-15 & 4.85159E-16  \\ 
2.01902 & 1.66693E-15 & 9.81417E-17 & 2.01902 & 2.16205E-15 & 1.27782E-16 & 2.01902 & 2.88466E-15 & 3.57436E-16  \\ 
2.02545 & 1.74980E-15 & 1.12876E-16 & 2.02545 & 2.13171E-15 & 1.25974E-16 & 2.02545 & 2.90724E-15 & 3.86474E-16  \\ 
2.03165 & 1.82837E-15 & 1.06535E-16 & 2.03165 & 2.08902E-15 & 1.16510E-16 & 2.03165 & 2.72256E-15 & 3.74589E-16  \\ 
2.0406  & 1.91342E-15 & 9.31512E-17 & 2.0406  & 2.17134E-15 & 1.19445E-16 & 2.0406  & 2.69971E-15 & 4.04758E-16  \\ 
2.04977 & 1.96417E-15 & 8.96671E-17 & 2.04977 & 2.32241E-15 & 1.05412E-16 & 2.04977 & 2.76381E-15 & 4.49176E-16  \\ 
2.05939 & 2.06224E-15 & 9.14667E-17 & 2.05939 & 2.34754E-15 & 9.66011E-17 & 2.05939 & 3.04182E-15 & 4.10525E-16  \\ 
2.06838 & 2.14385E-15 & 1.00008E-16 & 2.06838 & 2.45222E-15 & 1.06841E-16 & 2.06838 & 3.27945E-15 & 4.08582E-16  \\ 
2.07634 & 2.18938E-15 & 9.48017E-17 & 2.07634 & 2.56451E-15 & 1.11055E-16 & 2.07634 & 3.31831E-15 & 4.36232E-16  \\ 
2.08445 & 2.24628E-15 & 8.64849E-17 & 2.08445 & 2.65357E-15 & 1.06652E-16 & 2.08445 & 3.43399E-15 & 4.81935E-16  \\ 
2.09321 & 2.31436E-15 & 9.27182E-17 & 2.09321 & 2.68320E-15 & 1.11878E-16 & 2.09321 & 3.74805E-15 & 5.26672E-16  \\ 
2.10191 & 2.39196E-15 & 9.73509E-17 & 2.10191 & 2.72654E-15 & 1.09539E-16 & 2.10191 & 3.96360E-15 & 5.41321E-16  \\ 
2.11049 & 2.39048E-15 & 8.39771E-17 & 2.11049 & 2.71716E-15 & 9.62034E-17 & 2.11049 & 3.97294E-15 & 5.21006E-16  \\ 
2.11872 & 2.41553E-15 & 8.06201E-17 & 2.11872 & 2.72546E-15 & 1.04323E-16 & 2.11872 & 3.87817E-15 & 5.07542E-16  \\ 
2.12672 & 2.51533E-15 & 8.24227E-17 & 2.12672 & 2.86541E-15 & 1.20287E-16 & 2.12672 & 4.43579E-15 & 5.32689E-16  \\ 
2.13543 & 2.58310E-15 & 7.96663E-17 & 2.13543 & 2.92595E-15 & 1.16728E-16 & 2.13543 & 4.61147E-15 & 5.54098E-16  \\ 
2.1436  & 2.64299E-15 & 8.49349E-17 & 2.1436  & 2.93367E-15 & 1.19210E-16 & 2.1436  & 4.56206E-15 & 4.76897E-16  \\ 
2.15136 & 2.69969E-15 & 1.04414E-16 & 2.15136 & 2.89564E-15 & 1.35568E-16 & 2.15136 & 4.37690E-15 & 4.72580E-16  \\ 
2.15869 & 2.65026E-15 & 7.99967E-17 & 2.15869 & 2.71841E-15 & 1.10909E-16 & 2.15869 & 4.10472E-15 & 4.84592E-16  \\ 
2.16564 & 2.44528E-15 & 7.95661E-17 & 2.16564 & 2.46122E-15 & 1.15686E-16 & 2.16564 &  &   \\ 
2.17139 & 2.67017E-15 & 7.95072E-17 & 2.17139 & 2.53147E-15 & 1.59061E-16 & 2.17139 &  &   \\ 
2.17686 & 3.02815E-15 & 1.27952E-16 & 2.17686 & 2.50246E-15 & 2.02441E-16 & 2.17686 &  &   \\ 
2.18372 & 3.20672E-15 & 7.57003E-16 & 2.18372 & 2.92986E-15 & 7.19543E-16 & 2.18372 &  &   \\ 
2.11263 & 2.08220E-15 & 1.76446E-16 & 2.11263 & 2.05871E-15 & 2.52649E-16 & 2.11263 &  &   \\ 
2.12373 & 2.09741E-15 & 9.77512E-17 & 2.12373 & 2.26295E-15 & 1.73276E-16 & 2.12373 &  &   \\ 
2.12933 & 2.46793E-15 & 1.23434E-16 & 2.12933 & 2.63819E-15 & 1.51924E-16 & 2.12933 &  &   \\ 
2.13518 & 2.44913E-15 & 1.06935E-16 & 2.13518 & 2.67962E-15 & 1.36449E-16 & 2.13518 & 3.86352E-15 & 5.79448E-16  \\ 
2.14195 & 2.47287E-15 & 1.17533E-16 & 2.14195 & 2.69382E-15 & 1.26388E-16 & 2.14195 & 3.17669E-15 & 6.01422E-16  \\ 
2.14931 & 2.65851E-15 & 1.09262E-16 & 2.14931 & 2.81641E-15 & 1.18280E-16 & 2.14931 & 3.28187E-15 & 5.88484E-16  \\ 
2.15691 & 2.68898E-15 & 1.11227E-16 & 2.15691 & 2.71622E-15 & 1.15086E-16 & 2.15691 & 2.95467E-15 & 4.96899E-16  \\ 
2.16488 & 2.42844E-15 & 9.81085E-17 & 2.16488 & 2.48631E-15 & 1.16156E-16 & 2.16488 & 2.78672E-15 & 4.37103E-16  \\ 
2.17325 & 2.49722E-15 & 9.83552E-17 & 2.17325 & 2.72076E-15 & 1.14770E-16 & 2.17325 & 2.80869E-15 & 5.21278E-16  \\ 
2.18182 & 2.65795E-15 & 9.98303E-17 & 2.18182 & 2.99062E-15 & 1.08101E-16 & 2.18182 & 2.80975E-15 & 6.80945E-16  \\ 
2.18959 & 2.58605E-15 & 9.58642E-17 & 2.18959 & 2.88759E-15 & 1.08781E-16 & 2.18959 & 2.74018E-15 & 6.93414E-16  \\ 
2.19772 & 2.50921E-15 & 8.67402E-17 & 2.19772 & 3.05906E-15 & 1.20869E-16 & 2.19772 & 2.71717E-15 & 6.90955E-16  \\ 
2.20595 & 2.42639E-15 & 7.93267E-17 & 2.20595 & 3.17959E-15 & 1.37906E-16 & 2.20595 & 2.74455E-15 & 6.83315E-16  \\ 
2.21398 & 2.33961E-15 & 1.05765E-16 & 2.21398 & 3.06283E-15 & 1.45510E-16 & 2.21398 & 3.01822E-15 & 6.40675E-16  \\ 
2.22188 & 2.33618E-15 & 1.15906E-16 & 2.22188 & 3.19849E-15 & 1.46380E-16 & 2.22188 & 3.19992E-15 & 6.88742E-16  \\ 
2.22866 & 2.28167E-15 & 9.73196E-17 & 2.22866 & 3.20375E-15 & 1.49208E-16 & 2.22866 & 3.47627E-15 & 7.44844E-16  \\ 
2.23696 & 2.28333E-15 & 1.26116E-16 & 2.23696 & 3.24555E-15 & 1.37603E-16 & 2.23696 & 3.52536E-15 & 7.94122E-16  \\ 
2.24522 & 2.29734E-15 & 1.27417E-16 & 2.24522 & 3.01478E-15 & 1.57085E-16 & 2.24522 & 3.63362E-15 & 7.75792E-16  \\ 
2.2519  & 2.21698E-15 & 1.05537E-16 & 2.2519  & 2.82656E-15 & 1.58374E-16 & 2.2519  & 3.47115E-15 & 7.63492E-16  \\ 
2.25911 & 2.35405E-15 & 1.34450E-16 & 2.25911 & 3.07052E-15 & 1.62839E-16 & 2.25911 & 3.60626E-15 & 7.87863E-16  \\ 
2.26625 & 2.38989E-15 & 1.57034E-16 & 2.26625 & 3.12404E-15 & 1.67920E-16 & 2.26625 & 3.47341E-15 & 8.93141E-16  \\ 
2.2761  & 2.40367E-15 & 1.71859E-16 & 2.2761  & 3.07075E-15 & 2.27538E-16 & 2.2761  & 3.83657E-15 & 1.00173E-15  \\ 
2.28424 & 2.47536E-15 & 3.60119E-16 & 2.28424 & 3.28311E-15 & 3.82843E-16 & 2.28424 & 4.12671E-15 & 1.25140E-15  \\ 
2.29125 & 2.14047E-15 & 1.97341E-16 & 2.29125 & 2.91484E-15 & 2.28448E-16 & 2.29125 & 3.06066E-15 & 1.27180E-15  \\ 
2.29894 & 1.82729E-15 & 1.39034E-16 & 2.29894 & 2.54871E-15 & 2.13205E-16 & 2.29894 & 2.10342E-15 & 1.29480E-15  \\ 
2.30465 & 1.83450E-15 & 1.18788E-16 & 2.30465 & 3.00497E-15 & 2.56827E-16 & 2.30465 & 2.05041E-15 & 1.44621E-15  \\ 
2.31086 & 2.28488E-15 & 4.34866E-16 & 2.31086 & 3.18844E-15 & 4.97568E-16 & 2.31086 & 2.90437E-15 & 1.75422E-15  \\ 
2.31651 & 2.63187E-15 & 7.94890E-16 & 2.31651 & 3.11156E-15 & 8.20462E-16 & 2.31651 & 3.33332E-15 & 2.22156E-15  \\ 
2.32505 & 2.02439E-15 & 3.89874E-16 & 2.32505 & 2.75737E-15 & 4.14954E-16 & 2.32505 & 2.57146E-15 & 1.72444E-15  \\ 
2.32872 & 1.97924E-15 & 3.45239E-16 & 2.32872 & 2.94873E-15 & 3.64429E-16 & 2.32872 & 1.90773E-15 & 1.80550E-15  \\ 
2.33642 & 2.15101E-15 & 2.19251E-16 & 2.33642 & 2.39595E-15 & 2.87636E-16 & 2.33642 &  &   \\ 
2.34302 & 1.80835E-15 & 1.82389E-16 & 2.34302 & 2.08174E-15 & 3.85605E-16 & 2.34302 &  &   \\ 
2.35053 & 1.64353E-15 & 4.29150E-16 & 2.35053 & 2.58858E-15 & 5.47837E-16 & 2.35053 &  &   \\ 
2.35321 & 1.43145E-15 & 3.85427E-16 & 2.35321 & 2.88627E-15 & 5.00554E-16 & 2.35321 &  &   \\ 
2.36288 & 1.80437E-15 & 3.47187E-16 & 2.36288 & 3.09159E-15 & 4.73127E-16 & 2.36288 &  &   \\ 
2.36766 & 1.50262E-15 & 4.21286E-16 & 2.36766 & 2.59120E-15 & 5.11897E-16 & 2.36766 &  &   \\ 
2.36972 & 1.16197E-15 & 6.98525E-16 & 2.36972 & 2.02990E-15 & 9.53985E-16 & 2.36972 &  &   \\ 
\enddata
\end{deluxetable*}

\bibliographystyle{aasjournal}
\bibliography{references}

\clearpage

\end{document}